\documentclass%
[prd,twocolumn%
,a4paper,fleqn%
,nofootinbib,floatfix
,preprintnumbers
]{revtex4}%
\usepackage{amsfonts}
\usepackage{amsmath}
\usepackage{amssymb}
\usepackage{graphicx}
\usepackage{comment}
\usepackage{bm}
\usepackage{mathrsfs}
\usepackage{color}
\usepackage{ulem}
\providecommand{\U}[1]{\protect\rule{.1in}{.1in}}
\begin{document}
\preprint{CHIBA-EP-258}
\preprint{KEK Preprint 2023-24}
\title{Gauge-independent transition dividing the confinement phase in the lattice
SU(2) gauge-adjoint scalar model.}
\author{Akihiro Shibata}
\affiliation{Computing Research Center, High Energy Accelerator Research Organization
(KEK), Tsukuba 305-0801,Japan}
\affiliation{%
School of High Energy Accelerator Science,
SOKENDAI (The Graduate Univercity for
Advanced Studies), Tsukuba 305-0801, Japan}
\author{Kei-Ichi Kondo}
\affiliation{Department of Physics, Graduate School of Science, Chiba University, Chiba
263-8522, Japan}
\pacs{PACS number}

\begin{abstract}
The lattice SU(2) gauge-scalar model with the scalar field in the adjoint
representation of the gauge group has two completely separated confinement and
Higgs phases according to the preceding studies based on numerical simulations
that have been performed in the specific gauge fixing based on
 the conventional understanding of the Brout-Englert-Higgs mechanism.

In this paper, we reexamine this phase structure in a gauge-independent way
based on the numerical simulations performed without any gauge fixing. This is
motivated to confirm the recently proposed gauge-independent
Brout-Englert-Higgs mechanism for generating the mass of the gauge field
without relying on any spontaneous symmetry breaking. For this purpose, we
investigate  correlations between gauge-invariant operators obtained
by combining the original adjoint scalar field and the new field called
the color-direction field which is constructed from the gauge field based on 
the gauge-covariant decomposition of the gauge field due to Cho-Duan-Ge-Shabanov
and Faddeev-Niemi.

Consequently, we reproduce gauge independently the transition line separating
the confinement phase and the Higgs phase, and show surprisingly 
the existence of a new transition line that completely divides
 the confinement phase 
into two parts. Finally, we discuss the physical meaning of the new transition and
 the implications of the confinement mechanism.
\end{abstract}
\maketitle

\section{Introduction}

In this paper, we investigate the gauge-scalar model to clarify the mechanism
of confinement in the Yang-Mills theory in the presence of matter fields and
also nonperturbative characterization of the Brout-Englert-Higgs (BEH)
mechanism \cite{Higgs1} providing the gauge field with the mass, in the
gauge-independent way.
%In conventional studies on the lattice, gauge fixing %and color (global) symmetry fixing
%is necessary to obtain a signal due to Elizer's theorem.
%However, this does not satisfy the requirement of color confinement.
%To overcome this problem, we propose the lattice formulation of the gauge-independent
% Brout-Englert-Higgs (BEH) mechanism, which allows simulations and analysis without breaking the gauge symmetry.

For concreteness, we reexamine the lattice $SU(2)$ gauge-scalar model with a
radially fixed scalar field (no Higgs mode) which transforms according to the
adjoint representation of the gauge group $SU(2)$ without any gauge fixing. In
fact, this model was investigated long ago in \cite{Brower82} by taking a
specific gauge, say unitary gauge, based on the traditional characterization
for the BEH mechanism to identify the Higgs phase. It is a good place to
recall the traditional characterization of the BEH mechanism: If the original
continuous gauge group is spontaneously broken, the resulting massless
Nambu-Goldstone particle is absorbed into the gauge field to provide the gauge
field with the mass. In the perturbative treatment, such a spontaneous
symmetry breaking is signaled by the nonvanishing vacuum expectation value of
the scalar field. However, this is impossible to realize on the lattice unless
the gauge fixing condition is imposed, since gauge noninvariant operators
have vanishing vacuum expectation value on the lattice without gauge fixing
due to the Elitzur theorem \cite{Elitzur75}. This traditional characterization
of the BEH mechanism prevents us from investigating the Higgs phase in the
gauge-invariant way.

This  difficulty can be avoided by using the \textit{gauge-independent
description of the BEH mechanism} proposed recently by one of the authors
\cite{Kondo16,Kondo18}, which \textit{needs neither the spontaneous breaking
of gauge symmetry},
%$G \rightarrow H$,
nor the \textit{nonvanishing vacuum expectation value of the scalar field}.
%$\langle0|\phi(x)|0\rangle:=v\neq 0$.
Then we can give a \textit{gauge-invariant definition of the mass for the
gauge field} resulting from the BEH mechanism. Therefore, we can study the
Higgs phase in the gauge-invariant way on the lattice without gauge fixing
based on the lattice construction of gauge-independent description of the BEH
mechanism. Consequently, we can perform numerical simulations without any
gauge-fixing and compare our results with those of the preceding result
\cite{Brower82} obtained in a specific gauge. Indeed, our gauge-independent
study reproduces the transition line separating Higgs and confinement phases
obtained by \cite{Brower82} in a specific gauge.

%To explain it, we need to introduce a specific gauge-scalar model (\textit{complementary gauge-scalar model}) which reduces to the \textit{Yang-Mills theory 
%with a gauge-nvariant gluon mass term} (\textit{massive Yang-Mills theory}). The gauge-invariant gluon mass term simulates the dynamically generated mass to be extracted 
%in the low-energy effective theory of the Yang-Mills theory and plays the role of a new probe to study confinement mechanism through 
%the phase structure (Confinement, Higgs, deconfinement) in the gauge-invariant way. In this talk we give preliminary studies in this direction.%

Moreover, we investigate the phase structure of this model based on the
gauge-independent (invariant) procedure to look into the mechanism for
confinement. For this purpose we introduce the \textit{gauge-covariant decomposition
of the gauge field} originally due to Cho-Duan-Ge-Shabanov and Faddeev-Niemi
\cite{Cho80,Duan-Ge79,Shabanov99,FN98}, which we call CDGSFN decomposition for
short. It has been confirmed that this method is quite efficient to extract
the dominant mode responsible for quark confinement in a gauge-independent way
\cite{Exactdecomp09,CFNdccomp07,KKSS15}, even if we expect the dual superconductor picture for quark confinement \cite{dualsuper}.

To discriminate and characterize the phases among confinement phase, Higgs
phase, and the other possible phases, we investigate the correlations 
between the gauge-invariant composite operators constructed from the scalar
field and the color-direction field obtained through the CDGSFN
decomposition. As a result of the gauge-independent analysis, we find
surprisingly a new transition line that divides the conventional confinement
phase into two parts.
%Note that this finding owes much to gauge-independent simulations without any gauge fixing, and analysis and can only be established through this.
Finally, we discuss the physical meaning of this transition and the
implications to confinement.

This paper is organized as follows.
%In Sec.2 we give a brief review of the gauge-independent BEH mechanism for a SU(2) gauge-adjoint scalar model.
In Sec.II we define the lattice $SU(2)$ gauge-scalar model with a
radially fixed scalar field in the adjoint representation of the gauge group
and introduce the gauge-covariant CDGSFN decomposition of the gauge field
variable on the lattice. We explain the method of numerical simulations in the
new framework of the lattice gauge theory. In Sec.III we present the results
of the numerical simulations. We give an analysis in view of the the
gauge-covariant CDGSFN decomposition. By measuring the correlations 
between the gauge-invariant composite operators composed of the original
adjoint scalar field and the color-direction field obtained from the
decomposition, we find a new phase that divides the confinement phase
completely into two parts. 
We further examine two kinds of correlations  
between ($n \times n$)-Wilson loops $(n=1,2)$ to obtain the consistent result.
In Sec.IV, we discuss understanding the new phase structure obtained from numerical simulations.
The final section is devoted to conclusion and discussion. 
In the Appendix, 
we further examine the Wilson-loop operators 
to understand the transition between the two confinement phases (I) and (III).

\section{Lattice $SU(2)$ gauge-scalar model with a scalar field in the adjoint
representation}

%We now discuss the numerical simulations for the proposed gauge-scalar model on the lattice, where \textit{the gauge-independent description for the BEH mechanism}.%

\subsection{Lattice action} % and integration measure}

The $SU(2)$ gauge-scalar model with a radially fixed scalar field in the
adjoint representation is given on the lattice with a lattice spacing
$\epsilon$ by the following action with two parameters $\beta$ and $\gamma$:
\begin{align}
S_{\text{GS}}[U,\bm{\phi}] 
 &:=S_{G}[U]+S_{H}[U,\bm{\phi}]\,,\label{eq:GSmodel}\\
S_{G}[U] &:= \sum_{x, \mu<\nu}\frac{\beta}{2}\mathrm{tr}\left(
\bm{1} - U_{x,\mu\nu} \right) + \text{c.c.}
\notag\\
&= \sum_{x, \mu<\nu}\frac{\beta}{2}\mathrm{tr}
\left(
\bm{1}-U_{x,\mu}U_{x+\hat{\mu},\nu}U_{x+\hat{\nu},\mu}^{\dag}U_{x,\nu}^{\dag}
\right) \notag \\ 
& \quad \quad +\text{c.c.} \,, \label{eq:GaugeAction}\\
S_{H}[U,\bm{\phi}] 
  &   :=\sum_{x,\mu}\frac{\gamma}{2}\mathrm{tr}
   \left(  (D_{\mu}^{\epsilon}[U]\bm{\phi}_{x})^{\dag}(D_{\mu}^{\epsilon}[U]\bm{\phi}_{x})\right) \notag\\
 & =\sum_{x,\mu}{\gamma}\mathrm{tr}
     \left( \bm{1} - \bm{\phi}_{x} U_{x,\mu} \bm{\phi}_{x+\hat{\mu}}U_{x,\mu}^{\dagger}  \right) \,,
 \label{eq:ScalarAction}%
\end{align}
where $U_{x,\mu}=\exp(-ig\epsilon\mathscr{A}_{x,\mu}) \in SU(2)$ represents a
gauge variable on a link $\langle x,\mu\rangle$, 
$U_{x,\mu\nu}$ is the plaquette variable defined on a plaquette $\langle x,\mu\nu\rangle$ by
\begin{equation}
U_{x,\mu\nu}=U_{x,\mu}U_{x+\hat{\mu},\nu}%
U_{x+\hat{\nu},\mu}^{\dag}U_{x,\nu}^{\dag} ,
\end{equation}
\color{black}
$\bm{\phi}_{x}=%
\phi_{x}^{A}\sigma^{A}$ $\in su(2)$ ($A=1,2,3$) represents a scalar field on a
site $x$ in the adjoint representation subject to the radially fixed
condition: $\bm{\phi}_{x}\cdot\bm{\phi}_{x}=\phi_{x}^{A}\bm{\phi}_{x}%
^{A}=1$, and $D_{\mu}^{\epsilon}[U]\bm{\phi}_{x}$ represents the covariant
derivative in the adjoint representation defined as%
\begin{equation}
D_{\mu}^{\epsilon}[U]\bm{\phi}_{x}:=U_{x,\mu}\bm{\phi}_{x+\epsilon
\hat{\mu}}-\bm{\phi}_{x}U_{x,\mu}\text{ .} \label{eq:CovaritDriv}%
\end{equation}
This action reproduces in the naive continuum limit $\epsilon\to0$ the continuum
gauge-scalar theory with a radially fixed scalar field  $|\bm{\phi}_x|=v$ 
and a gauge coupling constant $g$ where $\beta=4/g^{2}$ and $\gamma=v^{2}/2$
\cite{Kondo16}.

The lattice action (\ref{eq:ScalarAction}) is invariant: 
$S_{\text{GS}}[U^\prime,\bm{\phi}^\prime] = S_{\text{GS}}[U,\bm{\phi}]$
under the local gauge transformation ${\Omega}_x \in \mathrm{SU(2)}_\mathrm{local}$
for the link variable $U_{x,\mu}$ and the site variable $\bm{\phi}_{x}$ given by 
\begin{align}
	& U_{x,\mu}
	\mapsto U_{x,\mu}^{\prime} = {\Omega}_x U_{x,\mu} {\Omega}_{x+\mu}^{\dagger} \, , \notag\\
	& \bm{\phi}_{x}  
	\mapsto \bm{\phi}^{\prime}_{x} = \Omega_x \bm{\phi}_x \Omega^{\dagger}_x \, . 
	\label{eq:gaugeTr}
\end{align}
Therefore, the color symmetry is preserved in the sense that $S_{\text{GS}}[U,\bm{\phi}]$ is invariant under the global transformation $\Omega_x = \Omega \in SU(2)_{\rm global}$ in Eq.(\ref{eq:gaugeTr}).

\subsection{Gauge-covariant decomposition}

To investigate gauge independently the phase structure of the gauge-scalar
model, we introduce the lattice version \cite{Exactdecomp09,CFNdccomp07} of
change of variables based on the idea of the \textit{gauge-covariant
decomposition of the gauge field}, so called the CDGSFN decomposition
\cite{Cho80,Duan-Ge79,Shabanov99,FN98}. For a review, see \cite{KKSS15}.

We introduce the (Lie algebra valued) site variable $\bm{n}_{x}:=n_{x}^{A}\sigma_{A}\in
su(2)-u(1)$ which is called the \textit{color-direction (vector) field}, in addition to
the original link variable $U_{x,\mu}\in SU(2)$. The link variable $U_{x,\mu}$
and the site variable $\bm{n}_{x}$ transform under the gauge
transformation $\Omega_{x}\in SU(2)$ as
\begin{equation}
U_{x,\mu}\rightarrow\Omega_{x}U_{x,\mu}\Omega_{x+\mu}^{\dagger}=U_{x,\mu
}^{\prime}, \quad \bm{n}_{x}\rightarrow\Omega_{x}\bm{n}_{x}\Omega
_{x}^{\dagger}=\bm{n}_{x}^{\prime}.
\end{equation}
In the decomposition, a link variable $U_{x,\mu}$ is decomposed into two
parts:
\begin{equation}
U_{x,\mu}:=X_{x,\mu}V_{x,\mu}.
\end{equation}
We identify the lattice variable $V_{x,\mu}$ with a link variable which
transforms in the same way as the original link variable $U_{x,\mu}$:
\begin{equation}
V_{x,\mu}\rightarrow\Omega_{x}V_{x,\mu}\Omega_{x+\mu}^{\dagger}=V_{x,\mu
}^{\prime}.
\end{equation}
On the other hand, we define the lattice variable $X_{x,\mu}$ such that it
transforms in just the same way as the site variable $\bm{n}_{x}$:
\begin{equation}
X_{x,\mu}\rightarrow\Omega_{x}X_{x,\mu}\Omega_{x}^{\dagger}=X_{x,\mu}^{\prime
},
\end{equation}
which automatically follows from the above definition of the decomposition.
Such decomposition is obtained by solving the defining equations:
\begin{align}
&  D_{\mu}[V]\bm{n}_{x}:=V_{x,\mu}\bm{n}_{x+\mu}-\bm{n}%
_{x}V_{x,\mu}=0,\\
&  \mathrm{tr}(\bm{n}_{x}X_{x,\mu})=0.
\end{align}
This defining equation has been solved exactly and the resulting link variable
$V_{x,\mu}$ and site variable $X_{x,\mu}$ are of the form
%(up to the normalization)
\cite{Exactdecomp09}:
\begin{align}
\tilde{V}_{x,\mu} & :=U_{x,\mu} +\bm{n}_{x}U_{x,\mu}\bm{n}_{x+\mu}, \notag \\
V_{x,\mu}  &  :=\tilde{V}_{x,\mu}/\sqrt{\mathrm{tr}(\tilde{V}_{x,\mu}^{\dagger}\tilde{V}_{x,\mu})/2}, \notag \\
X_{x,\mu}  &  :=U_{x,\mu}V_{x,\mu}^{-1}\text{ .}%
\label{eq:decomp}
\end{align}
This decomposition is obtained uniquely for a given set 
of link variable
 $U_{x,\mu}$ once the site variables $\bm{n}_{x}$ is given. The configurations of the
color-direction field $\{\bm{n}_{x}\}$ are obtained by minimizing the
functional:
\begin{equation}
F_{\text{red}}[\{\bm{n}_{x}\}| 
\{U_{x,\mu}\}]:=\sum_{x,\mu} \text{tr}%
\left\{  \left(  D_{x,\mu}[U]\bm{n}_{x}\right)  ^{\dag}\left(  D_{x,\mu
}[U]\bm{n}_{x}\right)  \right\}  , \label{reduction}%
\end{equation}
which we call the \textit{reduction condition}. 

Note that this functional has
the same form as the action of the scalar field:%
\begin{align}
S_{\phi}= \frac{\gamma}{2}F_{\text{red}}[\{\bm{\phi}_{x}\}|\{U_{x,\mu}\}] \,.
\end{align}
Therefore, the color-direction field obtained as the solution of the reduction condition satisfies the equations of motion of the adjoint scalar field.  In other words, the color-direction field is identified with degrees of freedom corresponding to the scalar field extracted from the gauge field.
Moreover, by using the color-direction fields, we can construct a gauge-invariant mass term  
without a scalar field \cite{Kondo16}.

\subsection{Numerical simulations}

The numerical simulation can be performed by updating link variables and scalar fields alternately. 
For the link variable $U_{x,\mu}$ we can apply the standard Hamiltonian Monte Carlo (HMC) algorithm, 
while for scalar field $\bm{\phi}_{x}$ $\in su(2)-u(1)$ 
we apply the adjoint-orbit representation for the reparametrization :%
%%%%%%%%%%%%%
\begin{equation}
\bm{\phi}_{x}:= Y_{x} \bm{\phi}_0  Y_{x}^{\dag}
              = Y_{x}  \sigma^3 Y_{x}^{\dag}
\,,  \quad  Y_{x} \in SU(2) ,
\label{eq:phi2}%
\end{equation}
%%%%%%%%%%%%%
which satisfies the normalization condition $\bm{\phi}_{x}\cdot \bm{\phi}_{x}=1$ automatically. 
Under the local gauge transformation ${\Omega}_x \in \mathrm{SU(2)}_\mathrm{local}$,
$Y_{x}$ is required to transform as
\begin{align} 
	& Y_{x} \mapsto Y_{x}^\prime = \Omega Y_{x} \,, 
\end{align}
to reproduce the transformation of $\bm{\phi}_{x}$: 
$\bm{\phi}_{x} \mapsto \bm{\phi}^{\prime}_{x} = \Omega_x \bm{\phi}_x \Omega^{\dagger}_x$.

Therefore,  we can replace the Haar measure $\prod_{x} d\bm{\phi}_x\delta(\bm{\phi}_{x}\cdot\bm{\phi}_{x}-1)$ by 
$\prod_{x} dY_{x}$, 
which enables us to apply the standard HMC algorithm for $Y_{x}$ 
to update configurations of the scalar fields $\bm{\phi}_{x}$.

\begin{figure}[hbt] \centering
	\includegraphics[height=71mm] {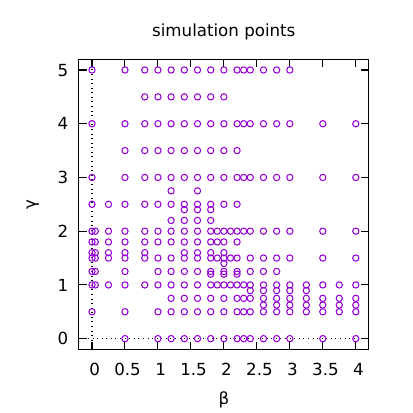}%
	\caption{
		Simulation points in $\beta$-$\gamma$ plane.
	}%
	\label{fig:simulation}%
\end{figure}

We perform Monte Carlo simulations on the $16^{4}$ lattice with periodic boundary condition in the gauge-independent way (without gauge fixing). 
In each Monte Carlo step (sweep), we update link variables $\{U_{x,\mu}\}$ and scalar fields $\{\bm{\phi}_{x}\}$ alternately by using the HMC algorithm with integral interval $\Delta\tau=1$ as explained in the previous section.
We take thermalization for $5000$ sweeps and store 800 configurations for measurements every 25 sweeps. 
Figure \ref{fig:simulation} shows datasets of the simulation parameters in the $\beta$--$\gamma$ plane.

\section{Lattice result and gauge-independent analyses }

\subsection{Action densities for the plaquette and scalar parts}%

The search for the phase boundary is performed by measuring the expectation value $\left\langle \mathcal{O} \right\rangle$ of a chosen operator $\mathcal{O}$ by changing $\gamma$ (or $\beta$) along the $\beta= \text{const.}$  (or $\gamma=\text{const.}$) lines. 
In order to identify the boundary, we used the bent, step, and gap observed in the graph of the plots for $\left\langle \mathcal{O} \right\rangle$. 
%The existence and the magnitude of the gap is to be related to the existence and the strength of the first-order phase transition.

First of all, in order to determine the phase boundary of the model, we
measure the Wilson action per plaquette (plaquette-action density), 
\begin{align}
P &:=\frac{1}{6N_{\text{site}}}\sum_{x,\mu<\nu}\frac{1}{2}\mathrm{tr}%
(U_{x,\mu\nu})\,, \label{eq:P}%
\end{align}
%with  
%$U_{x,\mu\nu}=U_{x,\mu}U_{x+\hat{\mu},\nu}%
%U_{x+\hat{\nu},\mu}^{\dag}U_{x,\nu}^{\dag}$,
and 
the scalar action per link (scalar-action density),
\begin{equation}
M:=\frac{1}{4N_{\text{site}}}\sum_{x,\mu}\frac{1}{2}\mathrm{tr}\left(
\left(  D_{\mu}[U_{x,\mu}]\bm{\phi}_{x}\right)  ^{\dag}\left(  D_{\mu
}[U_{x,\mu}]\bm{\phi}_{x}\right)  \right) \,, \label{eq:S}%
\end{equation}
as Brower et al. have done in \cite{Brower82}. \ %

\begin{figure*}[hbt] \centering
\includegraphics[height=57mm] {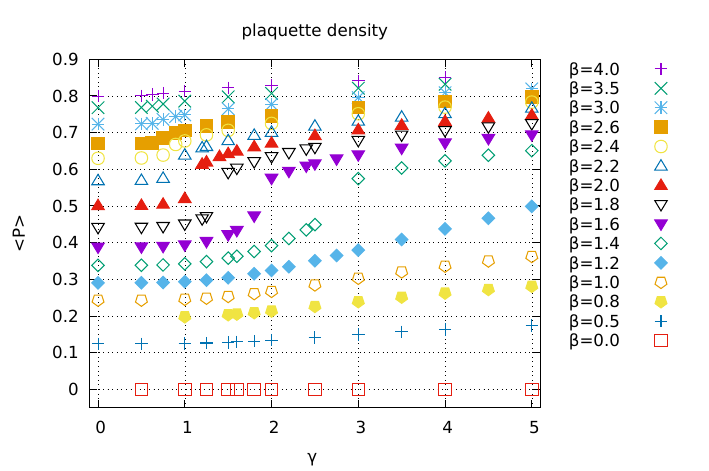} 
\includegraphics[height=57mm] {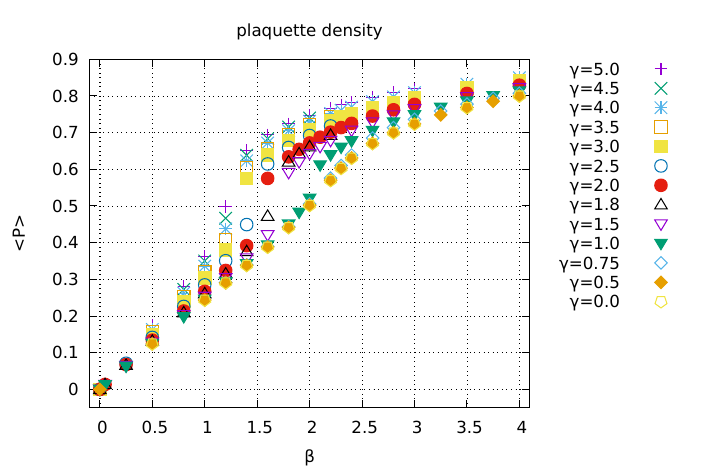}%
%EndExpansion%
\caption{
Average of the plaquette-action density $\left\langle P \right\rangle$.
Left: $\left\langle P \right\rangle$ versus $\gamma$ on various $\beta=$ const. lines. Right: $\left\langle P \right\rangle$ versus $\beta$ on  various $\gamma=$const. lines. 
}%
\label{fig:mesurement<P>}%
\end{figure*}%
\begin{figure*}[hbt] \centering
\includegraphics[height=57mm] {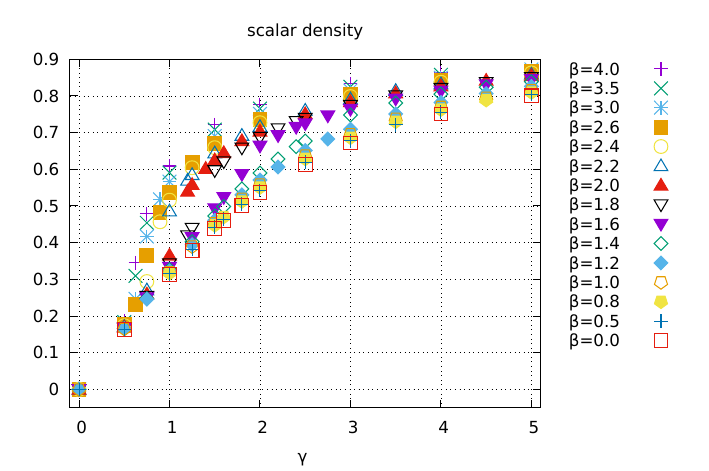} 
\includegraphics[height=57mm] {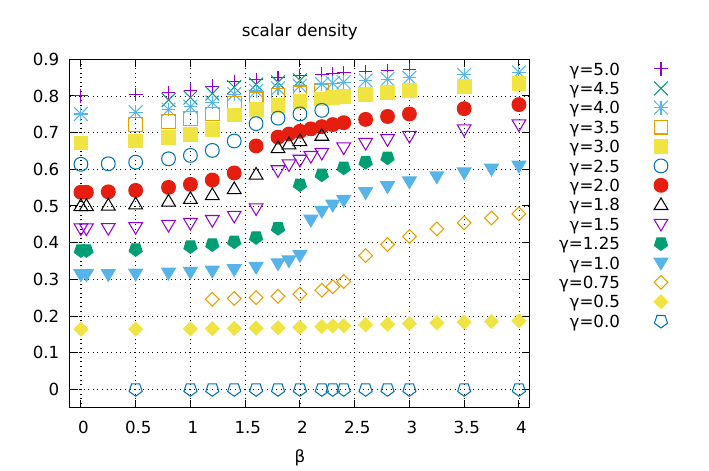} 
\caption{
Average of the scalar-action density $\left\langle M \right\rangle$.
Left: $\left\langle M \right\rangle$ versus $\gamma$ on various $\beta=$ const. lines.
Right: $\left\langle M \right\rangle$ versus $\beta$ on various $\gamma=$const. lines. 
}%
\label{fig:mesurement<M>}%
\end{figure*}%

First, we try to determine the phase boundary from the plaquette-action density. 
Figure \ref{fig:mesurement<P>} shows the results of measurements of the plaquette-action density
$\left\langle P\right\rangle $ in the $\beta$--$\gamma$ plane. 
The left panel shows the plots of $\left\langle P\right\rangle $\ along $\beta=\text{const.}$ lines as functions of $\gamma$, where error bars are not shown because they are smaller than the size of the plot points. 
On the other hand, the right panel shows the plots of $\left\langle P\right\rangle $\ along $\gamma=\text{const.}$ lines as functions of $\beta$.

Next, in the same way, we try to determine the phase boundary from the scalar-action density. 
Figure \ref{fig:mesurement<M>} shows the results of measurement of the scalar-action density $\left\langle M\right\rangle $ in the $\beta$--$\gamma$ plane. 
The left panel of Fig.\ref{fig:mesurement<M>} shows the plots of $\left\langle M\right\rangle $ along  $\beta=\text{const.}$ lines as functions of $\gamma$,
while the right panel of Fig.\ref{fig:mesurement<M>} shows the plots of $\left\langle M\right\rangle $ along $\gamma=\text{const.}$ lines as functions of $\beta$.

\begin{figure*}[hbt] \centering
\includegraphics[height=63mm] {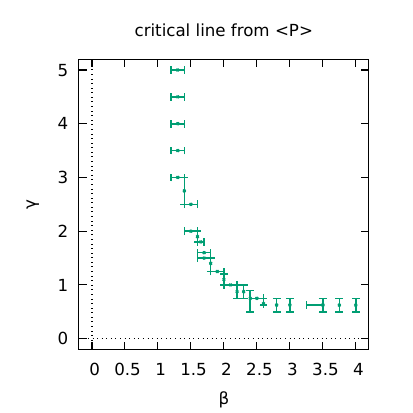} \quad \quad
\includegraphics[height=63mm ]{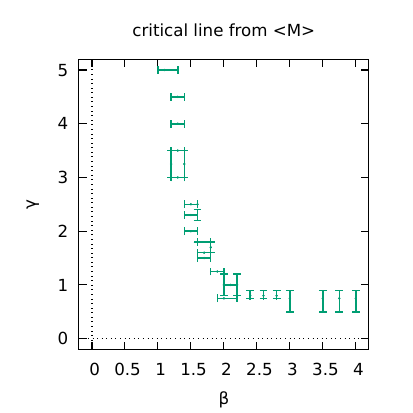}%
\caption{
The phase boundary determined by the action densities.
Left: $\left\langle P \right\rangle$.
Right: $\left\langle M \right\rangle$.
}%
\label{fig:Critical_action}%
\end{figure*}

Figure \ref{fig:Critical_action} shows the resulting phase boundary.
The left panel presents the phase boundary determined from the plaquette-action density $\left\langle P\right\rangle$ (see Fig.\ref{fig:mesurement<P>}).
The right panel shows the phase boundary determined from the scalar-action density $\left\langle M\right\rangle$ (see Fig.\ref{fig:mesurement<M>}).
%The right panel of Figure \ref{fig:Critical_action} shows the phase boundary determined from scalar-action density. 
The interval between the two simulation points corresponds to the short line with ends. 
The error bars in the phase boundary are due to the spacing of the simulation points.
It should be noticed that the two phase boundaries determined from 
$\left\langle P\right\rangle$ and $\left\langle M\right\rangle$ are consistent 
within accuracy of numerical calculations. 
Thus we find that the gauge-independent numerical simulations reproduce 
the phase boundary  obtained by Brower {\it et al}. \cite{Brower82}.%

\subsection{Susceptibilities for $P$ and $M$}%
\label{sec:sus_P&M}
To find out more about phase boundary, we next measure \textquotedblleft susceptibility" for the action densities:%
\begin{align}
\left\langle \chi(P)\right\rangle  &  := 
(6N_{\text{site}}) \left[ \left\langle P^{2}\right\rangle
-\left\langle P\right\rangle ^{2} \right] \text{ ,}\label{eq:Chi(P)}\\
\left\langle \chi(M)\right\rangle  &  := 
(4N_{\text{site}}) \left[ \left\langle M^{2}\right\rangle
-\left\langle M\right\rangle^{2} \right] \text{ .}\label{eq:Chi(S)}%
\end{align}
\begin{figure*}[hbt] \centering
\includegraphics[height=54mm] {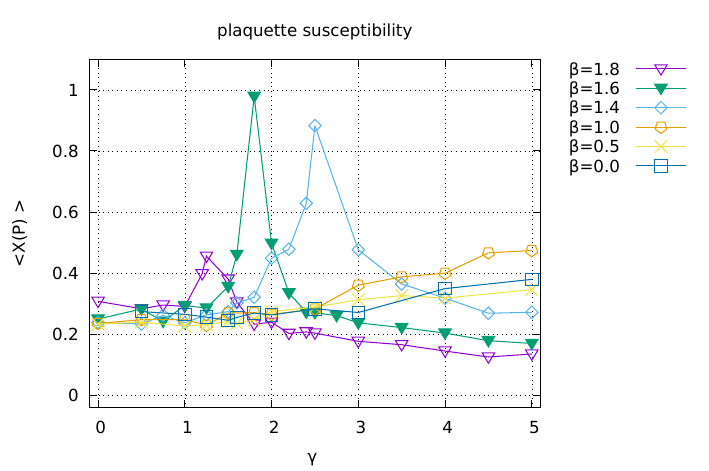}
\includegraphics[height=54mm] {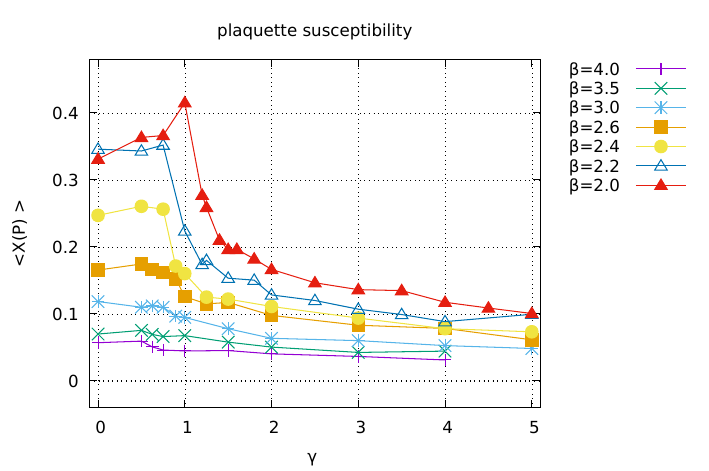}
\includegraphics[height=54mm] {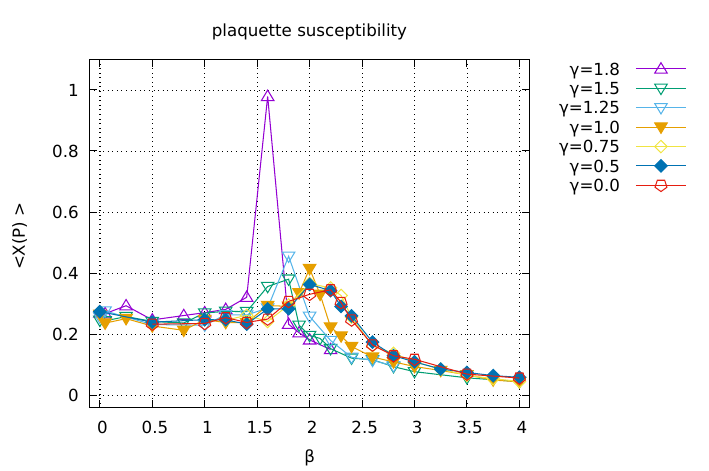}
\includegraphics[height=54mm] {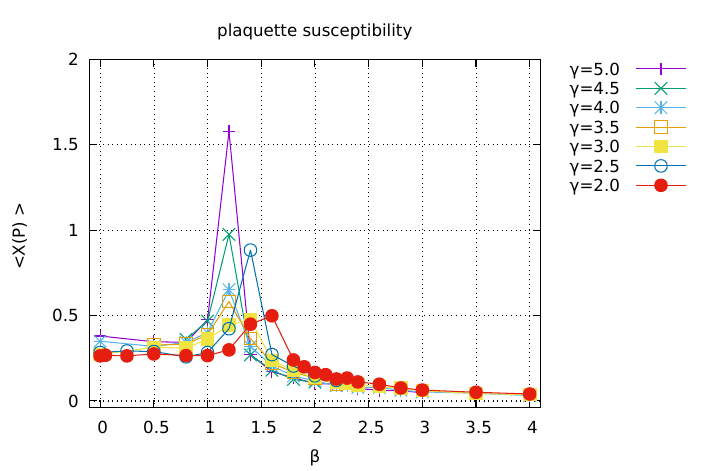}
\caption{
Susceptibility of plaquette $\left\langle \chi(P)\right\rangle$.
Upper panels: $\left\langle \chi(P)\right\rangle$ versus $\gamma$ on  various $\beta=$const. lines.
Lower panels: $\left\langle \chi(P)\right\rangle$ versus $\beta$ on various $\gamma=$const. lines.
}%
\label{fig:susceptiblity<P>}
\end{figure*}%
\begin{figure*}[hbt] \centering
\includegraphics[height=54mm] {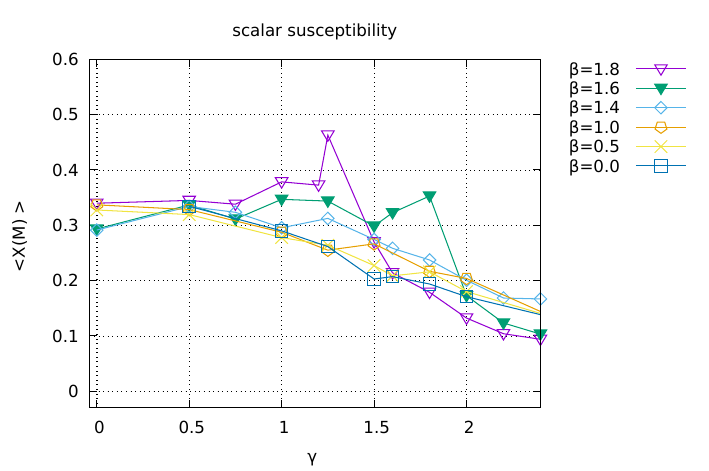}
\includegraphics[height=54mm] {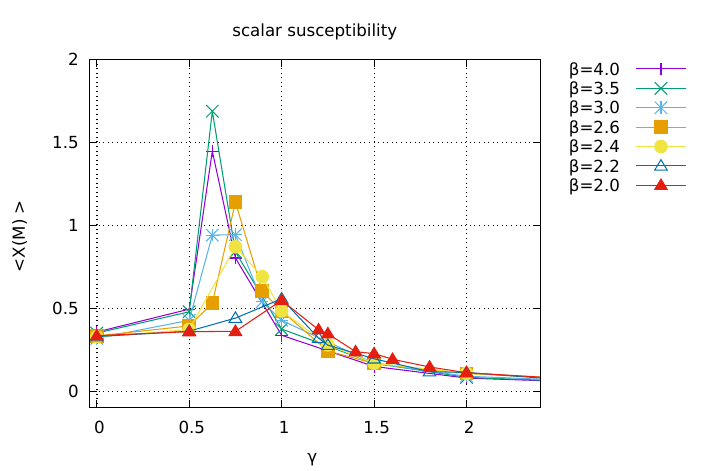}
\includegraphics[height=54mm] {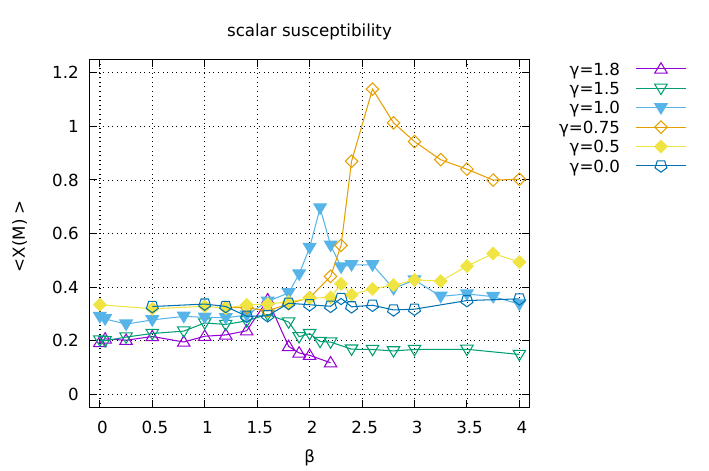}
\includegraphics[height=54mm] {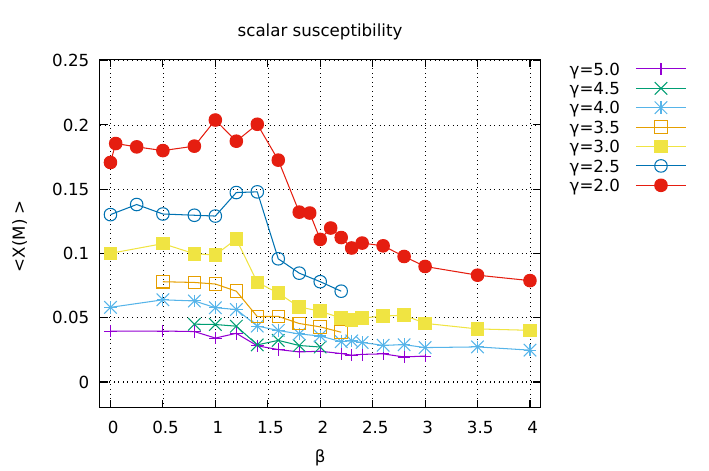}
\caption{
Susceptibility for the scalar action $\left\langle \chi(M)\right\rangle$.
Upper panels: $\left\langle \chi(M)\right\rangle$ versus $\gamma$ on  various $\beta=$const. lines.
Lower panels: $\left\langle \chi(M)\right\rangle$ versus $\beta$ on various $\gamma=$const. lines.}
\label{fig:susceptibility<M>}%
\end{figure*}%

Figure \ref{fig:susceptiblity<P>} shows the measurements of $\left\langle
\chi(P)\right\rangle$. 
The upper panels show plots of $\left\langle \chi(P)\right\rangle $ versus $\gamma$ along $\beta=$ const. lines, while the lower panels show plots of $\left\langle \chi(P)\right\rangle $  versus $\beta$ along the $\gamma=$ const. lines. 

%We find that some plots of  $\left\langle \chi(P)\right\rangle $ indicate peaks or blowup shapes, and plot these location in $\beta$--$\gamma$ plane as the left panel of Figure \ref{fig:criticalChi(P)VChi(M)>}. 

Figure \ref{fig:susceptibility<M>} shows the measurements of $\left\langle \chi(M)\right\rangle$. 
The upper panels show plots of $\left\langle \chi(M)\right\rangle $ versus $\gamma$ along $\beta=$ const. lines, 
while the lower panels show plots of $\left\langle \chi(M)\right\rangle $  versus $\beta$ along the $\gamma=$ const. lines.

\begin{figure*}[hbt] \centering
\includegraphics[height=63mm, trim= 20 0 20 0 ,clip ]
{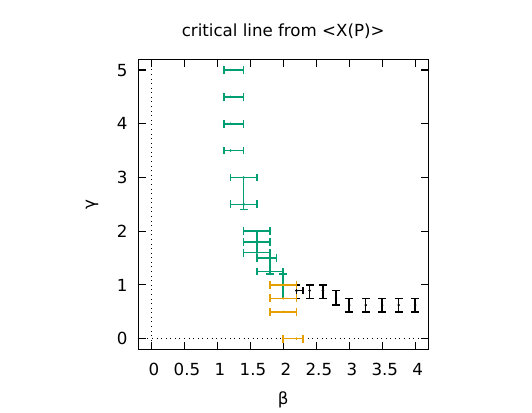}
\quad
\includegraphics[height=63mm, trim= 20 0 20 0 , clip]
{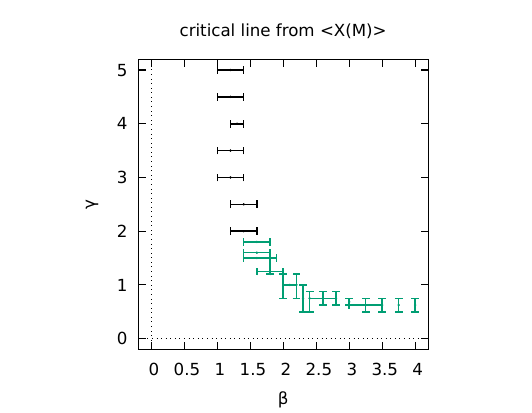}
\caption{
Critical lines determined from susceptibilities.
Left: from $\left\langle \chi(P) \right\rangle$.
Right: from $\left\langle \chi(M) \right\rangle$.
}%
\label{fig:criticalChi(P)VChi(M)>}%
\end{figure*}%

Figure \ref{fig:criticalChi(P)VChi(M)>} shows the phase boundary determined by the susceptibility (specific heat) as a function of $\beta$ or $\gamma$. The green boundary is determined from the position of the peak in the susceptibility graph. 
The black boundary was determined from the position of the bend in the susceptibility graph.
The orange boundary in the left panel of {Fig.\ref{fig:criticalChi(P)VChi(M)>}}
is determined from the peak position of the plaquette-action susceptibility.

The left panel of Fig.\ref{fig:criticalChi(P)VChi(M)>} gives same boundary as that determined by $\left\langle \chi(P)\right\rangle $ in Fig.\ref{fig:Critical_action} for relatively large $\gamma$.

It should be remarked that the phase boundaries in Fig.\ref{fig:criticalChi(P)VChi(M)>} and Fig.\ref{fig:Critical_action} do not necessarily coincide: 
in Fig.\ref{fig:Critical_action} the boundary line in the region $\beta>2$ (the black line) extends along the horizontal line $\gamma\simeq1$ toward the pure scalar axis at $\beta=\infty$, while in the left panel of Fig.\ref{fig:criticalChi(P)VChi(M)>} the boundary line extends also to the point $(\beta\simeq2.2,\gamma=0)$ on the pure gauge axis at $\gamma=0$ (the orange line).

However, the orange part of the phase boundary can only be found from the measurement of susceptibility $\left\langle \chi(P) \right\rangle$. Indeed, it cannot be found from the measurement of $\left\langle \chi(M) \right\rangle$. This fact indicates that the orange part is not the phase boundary and could be the crossover, discriminating the weak coupling asymptotic scaling region from the strong coupling region, as seen by Bhanot and Creutz in their model \cite{BhanotCreutz81}.

The right panel of Fig.\ref{fig:criticalChi(P)VChi(M)>} shows the phase boundary determined by  measurements of $\left\langle \chi (M)\right\rangle $ given in Fig.\ref{fig:susceptibility<M>}. 
The phase boundary determined from $\left\langle \chi(P)\right\rangle$ and $\left\langle \chi(M)\right\rangle$ coincide. 
Note that the location of the phase boundary is well approximated by the two lines: the right part of $\gamma \simeq 0.75$ and the upper part of $\beta \simeq 1.2$ except for the neighborhood of their intersection point.
%This indicates the nature of the first-order phase transition.%

%%%%%%%%%%%%%%%%%%%%%%%%%%%%%%%%%%%%%%%%%%%%%%%%%%
%%%% Figure for
%%%% Correlations between the scalar field and the color-direction field through the gauge covariant decompositionA
%%%%%%%%%%%%%%%%%%%%%%%%%%%%%%%%%%%%%%%%%%%%%%%%%%%%%%%%%%%%%%%%
\begin{figure*}[hbt] \centering
	\includegraphics[height=55mm] {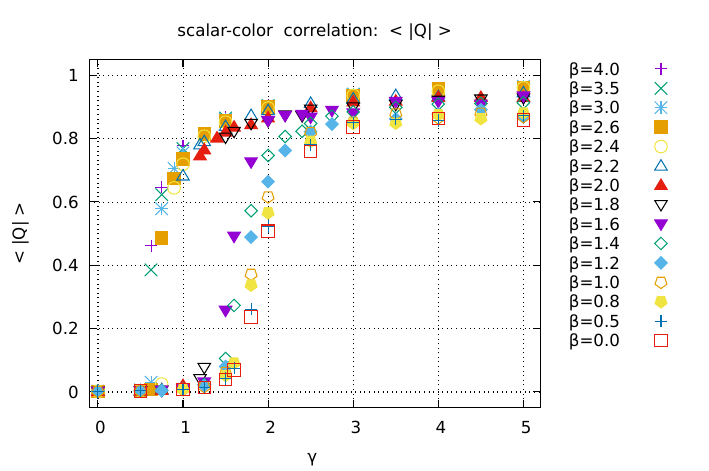}
	\includegraphics[height=55mm] {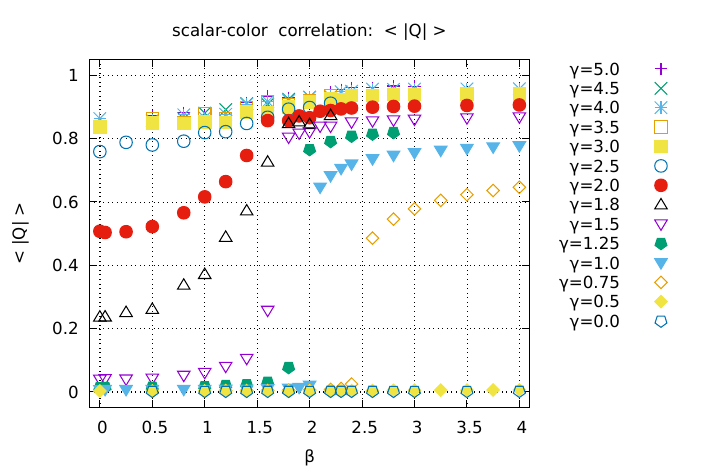}%
	\caption{
		Average of the scalar-color composite field $\left\langle \left\vert Q\right\vert \right\rangle$.
		Left: $\left\langle \left\vert Q\right\vert \right\rangle$ versus $\gamma$ on various $\beta= \text{const.}$ lines. 
		Right: $\left\langle \left\vert Q\right\vert \right\rangle$ versus $\beta$ on various $\gamma=\text{const.}$ lines. 
	}%
	\label{fig:measurement<Q>}%
\end{figure*}%

%%%%%%%%%%%%%%%%%%%%%%%%%%%%%
\begin{figure*}[hbt] \centering
	\includegraphics [height=55mm] {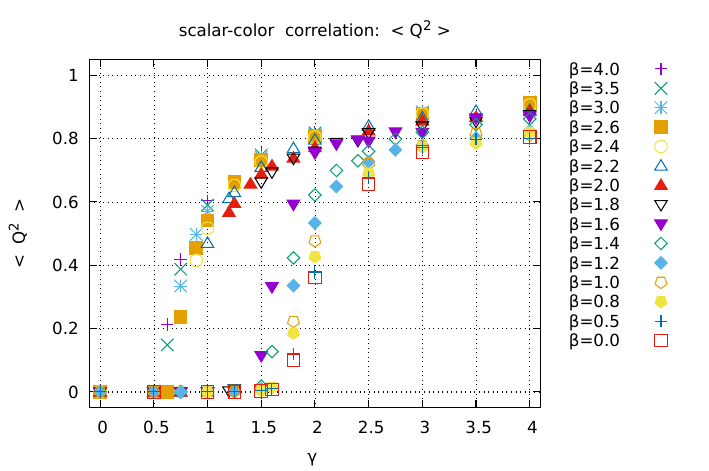} 
	\includegraphics [height=55mm] {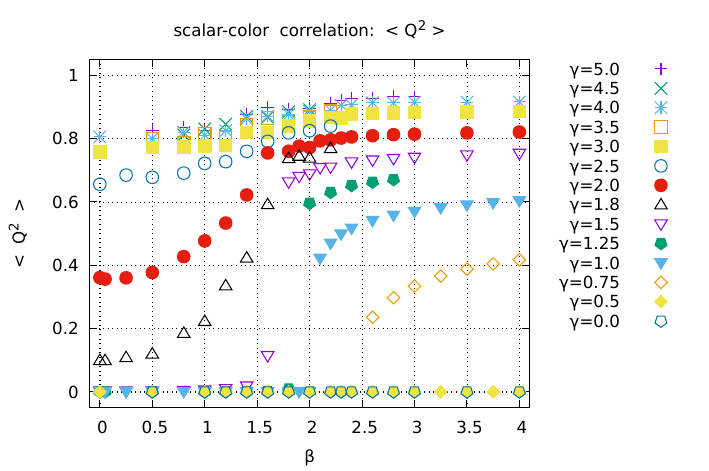}%
	\caption{
		Average of the squared scalar-color composite field $\left\langle Q^{2}\right\rangle$.
		Left: $\left\langle Q^{2}\right\rangle$ versus $\gamma$ on various $\beta=\text{const.}$ lines. 
		Right: $\left\langle Q^{2}\right\rangle$ versus $\beta$ on various $\gamma=\text{const.}$ lines. 
	}%
	\label{fig:measurement<Q2>}%
\end{figure*}%

%%%%%%%%%%%%%%%

%%%%%%%%%%%%%%%%%%%%%%%%%%%%%%%%%%%%%%%%%%%%%%%%%%%%%%%%%%%%%%%%%%%%%%%%%%%%%%%%%%%%%%%%%%%%%%%%%%%%%%%%%%%%%%%%%%%%

%\subsection{Analysis in view of the gauge covariant decomposition}

\subsection{Correlations between the scalar field and the color-direction field through the gauge covariant decomposition}%

We measure the scalar-color correlation detected by the scalar-color composite operator:%
\begin{equation}
Q=\frac{1}{N_{\text{site}}}\sum_{x}\frac{1}{2}\mathrm{tr}(\bm{n}%
_{x}\bm{\phi}_{x}), \label{eq:Q}
\end{equation}
%\ref{fig:critical<Q>}
where $\bm{n}_{x}$ is the color-direction field in the gauge-covariant decomposition for the gauge link variable. 
For this purpose, we need  to solve the reduction condition (\ref{reduction}) to obtain the color-direction field $\bm{n}_{x}$, which however has two kinds of ambiguity. 
One comes from so-called the Gribov copies that are  the local minimal solutions of the reduction condition.  
In order to avoid the local minimal solutions and to obtain the absolute minima, the reduction condition is solved by changing the initial values to search for the absolute minima of the functional. 
The other comes from the choice of a global sign factor, which originates from the fact that whenever a configuration $\{\bm{n}_{x}\}$ is a solution, the flipped one $\{-\bm{n}_{x}\}$ is also a solution, since the reduction functional is quadratic in the color-direction fields. 
To avoid these issues, we propose to use $\left\langle \left\vert Q\right\vert \right\rangle$ and $\left\langle Q^{2}\right\rangle$, which are examined as the order parameters that determine the phase boundary.%

The phase boundary is searched for based on two ways:

\begin{description}
\item[(i)] 
the location at which $\left\langle \left\vert Q\right\vert
\right\rangle $ changes from $\left\langle \left\vert Q\right\vert
\right\rangle \simeq0$ to $\left\langle \left\vert Q\right\vert \right\rangle
>0$.
This is also the case for $\left\langle Q^{2}\right\rangle$.

\item[(ii)] 
the location at which $\left\langle \left\vert
Q\right\vert \right\rangle $ changes abruptly, as was done for
$\left\langle P\right\rangle $ and $\left\langle M\right\rangle$.
This is also the case for $\left\langle Q^{2}\right\rangle$.

\end{description}

Figure \ref{fig:measurement<Q>} shows the measurements of $\left\langle \left\vert Q\right\vert \right\rangle$. The left panel shows plots of $\left\langle \left\vert Q\right\vert \right\rangle $ versus $\gamma$ along various $\beta=\text{const.}$ lines, while the right panel shows plots of $\left\langle \left\vert Q\right\vert \right\rangle $ versus $\beta$ along various $\gamma=\text{const.}$ lines. 

Figure \ref{fig:measurement<Q2>}, on the other hand, shows the measurements of
$\left\langle Q^{2}\right\rangle $ in the same manner as $\left\langle \left\vert Q\right\vert \right\rangle $. 

%%%%%%%%%%%%%%%%%%%
\begin{figure*}[hbt] \centering
\includegraphics[height=63mm] {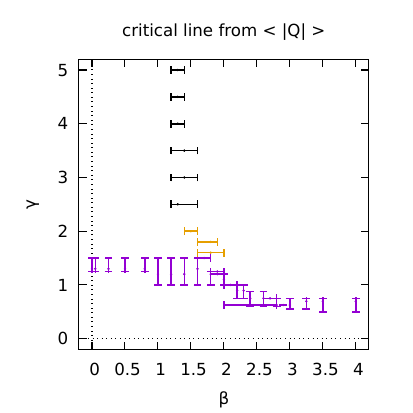} \quad \quad
\includegraphics[height=63mm] {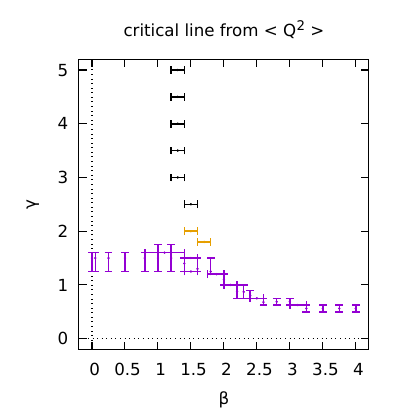}
\caption{
Critical lines determined (Left) from $\left\langle \left\vert Q\right\vert \right\rangle$, (Right) from $\left\langle Q^{2}\right\rangle$.
}%
\label{fig:critical<Q>}
\end{figure*}%
%%%%%%%%%%%%%%%%

Figure \ref{fig:critical<Q>} shows the phase boundary (critical line) determined by
$\left\langle \left\vert Q\right\vert \right\rangle $ and $\left\langle Q^{2}\right\rangle$.
The left panel of Fig.\ref{fig:critical<Q>} shows the phase boundary determined from 
$\left\langle \left\vert Q\right\vert \right\rangle $. 
The right panel of Fig.\ref{fig:critical<Q>} shows the phase boundary determined from $\left\langle Q^{2}\right\rangle $.

The purple boundary indicates that (i) $\left\langle \left\vert Q\right\vert \right\rangle $ changes from $\left\langle \left\vert Q\right\vert \right\rangle \simeq0$ to $\left\langle \left\vert Q\right\vert \right\rangle >0$ 
(or $\left\langle Q^{2}\right\rangle $ changes from $\left\langle Q^{2}\right\rangle \simeq0$ to $\left\langle Q^{2}\right\rangle >0$). 
The black boundary corresponds to the location  at which $\left\langle \left\vert Q\right\vert \right\rangle $ (or $\left\langle Q^{2}\right\rangle $) has gaps. 
The orange boundary corresponds to the location at which $\left\langle \left\vert Q\right\vert \right\rangle $ (or $\left\langle Q^{2}\right\rangle $) bends.
The results of left and right panels in Fig.\ref{fig:critical<Q>} are consistent with each other.

Figure \ref{fig:critical<Q>} shows not only the phase boundary that divides the phase diagram into two phases, the so-called Higgs phase and the confinement phase, but also the new boundary that divides the confinement phase into two different parts. 
It should be remarked that this finding owes much to gauge-independent numerical simulations and their analyses, and these new results can only be established through our framework.

%%%%%%%%%%%%%%%%%%%%%%%%%%%%%%%%
\begin{figure*}[hbt] \centering
\includegraphics[height=55mm] {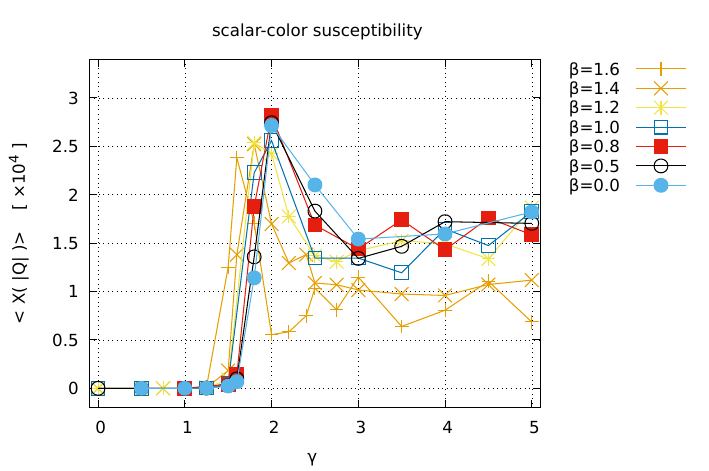}
\includegraphics[height=55mm] {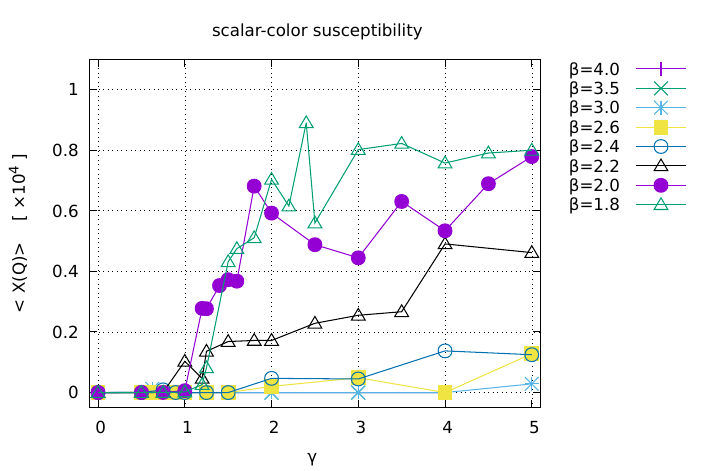}
\includegraphics[height=55mm] {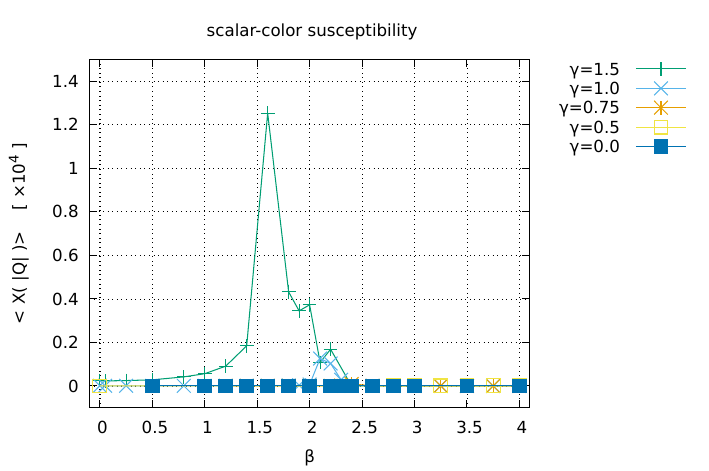}
\includegraphics[height=55mm] {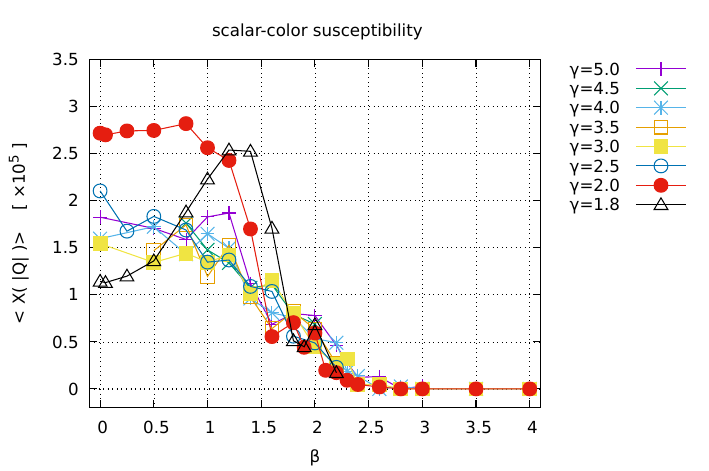}

\caption{
Susceptibility for the scalar-color field $\left\langle \chi(\left\vert Q\right\vert )\right\rangle$.
Upper panels: $\left\langle \chi(\left\vert Q\right\vert )\right\rangle$ versus $\gamma$ on  various $\beta=$const. lines.
Lower panels: $\left\langle \chi(\left\vert Q\right\vert )\right\rangle$ versus $\beta$ on various $\gamma=$const. lines. 
}%
\label{fig:susceptibility<Q>}%
\end{figure*}%

%%%%%%%%%%%%%%%%%%%%%%%%%%%%%%%%%%
\begin{figure}[hbt] \centering
	\includegraphics[height=65mm] {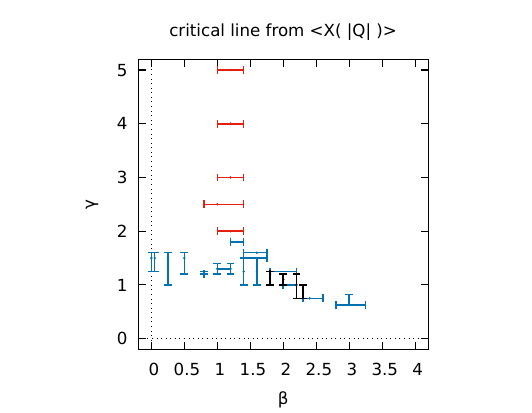}%
	%EndExpansion%
	\caption{
		Critical line determined from the susceptibility $\left\langle \chi(\left\vert Q\right\vert )\right\rangle$.
	}
	\label{fig:critical<Chi(|Q|)>}%
\end{figure}%

\subsection{Susceptibility for the scalar and color-direction field  }%

Next we investigate the ``susceptibility" of the scalar-color local
correlation:%
\begin{equation}
\left\langle \chi(\left\vert Q\right\vert )\right\rangle =
(4N_{\text{site}})\left[
\left\langle
Q^{2}\right\rangle -\left\langle \left\vert Q\right\vert \right\rangle ^{2}%
\right]
\end{equation}

Figure \ref{fig:susceptibility<Q>} shows the measurement of $\left\langle \chi(\left\vert Q\right\vert )\right\rangle$. 
The upper panels of Fig.\ref{fig:susceptibility<Q>} show plots of $\left\langle \chi(\left\vert Q\right\vert )\right\rangle $ versus $\gamma$ along the $\beta=\text{const.}$ lines, while   the lower panels show plots of $\left\langle \chi(\left\vert Q\right\vert )\right\rangle $ versus $\beta$  along the $\gamma=\text{const.}$ lines.

First, we search for the transition along the vertical lines with fixed values of $\beta$ in a phase diagram. 
For relatively small fixed value of $\beta$ ($0\le \beta \le 1.6$), $\left\langle \chi(\left\vert Q\right\vert )\right\rangle$ is nearly equal to zero for small $\gamma$, but reaches a large but finite value across a critical value $\gamma_c(\beta)$, showing a peak as $\gamma$ increases. 
For larger values of $\beta$ ($1.8\le \beta \le 4.0$), $\left\langle \chi(\left\vert Q\right\vert )\right\rangle$ increases from zero to a finite value, which shows however no peak and increases monotonically as $\gamma$ increase. 
These observations yield the existence of a new transition line $\gamma=\gamma_c(\beta)$.
%%, which seems to be the first order transition line. 

Next, we search for the transition along the horizontal line  with fixed values of $\gamma$ in a phase diagram. 
For small fixed value of $\gamma$ ($0\le \gamma \le 1.5$), $\left\langle \chi(\left\vert Q\right\vert )\right\rangle$ shows nonzero value for small $\beta$, and decreases monotonically as $\beta$ increases. 

Figure \ref{fig:critical<Chi(|Q|)>} shows the phase boundary (critical line) determined from $\left\langle \chi(\left\vert Q\right\vert )\right\rangle$.
 The blue boundary is obtained from the the location of the rapid change. 
 The red and black intervals are obtained from the location of bends.
This result agrees with the critical line already obtained above.

\begin{figure}[hbt] \centering
	\includegraphics[height=35mm]{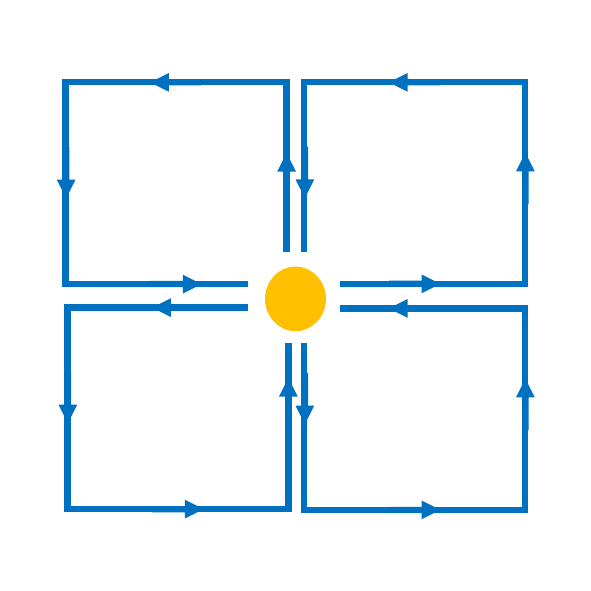}
	\quad 
	\includegraphics[height=34mm]{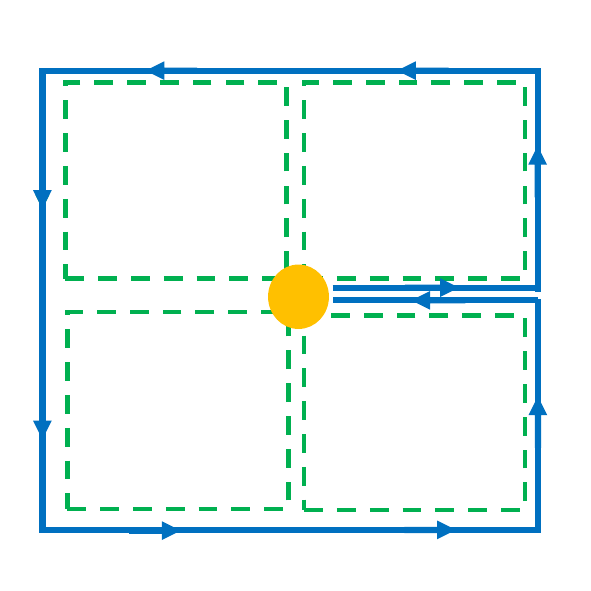}
	\caption{ Left: the graphical representation of Eq.(\ref{eq:hatR}). 
		The green lines represent link variables $U_{x,\mu}$ and the orange circle represents the position where the scalar field  $\boldsymbol{\phi}_{x}$ is inserted. 
		Right: the graphical representation of $\widehat{W}^{(2\times 2)}_{x,\mu\nu}$ in Eq.(\ref{eq:hatW}). A Wilson loop of the 2$\times$2 rectangular is connected by the Schwinger lines. The orange circle represents the position where the scalar field is inserted. }
	\label{fig:measurement-RW}
\end{figure}%

%%%%%%%%%%%%%
\begin{figure*}[hbt] \centering
	\includegraphics[height=60mm]{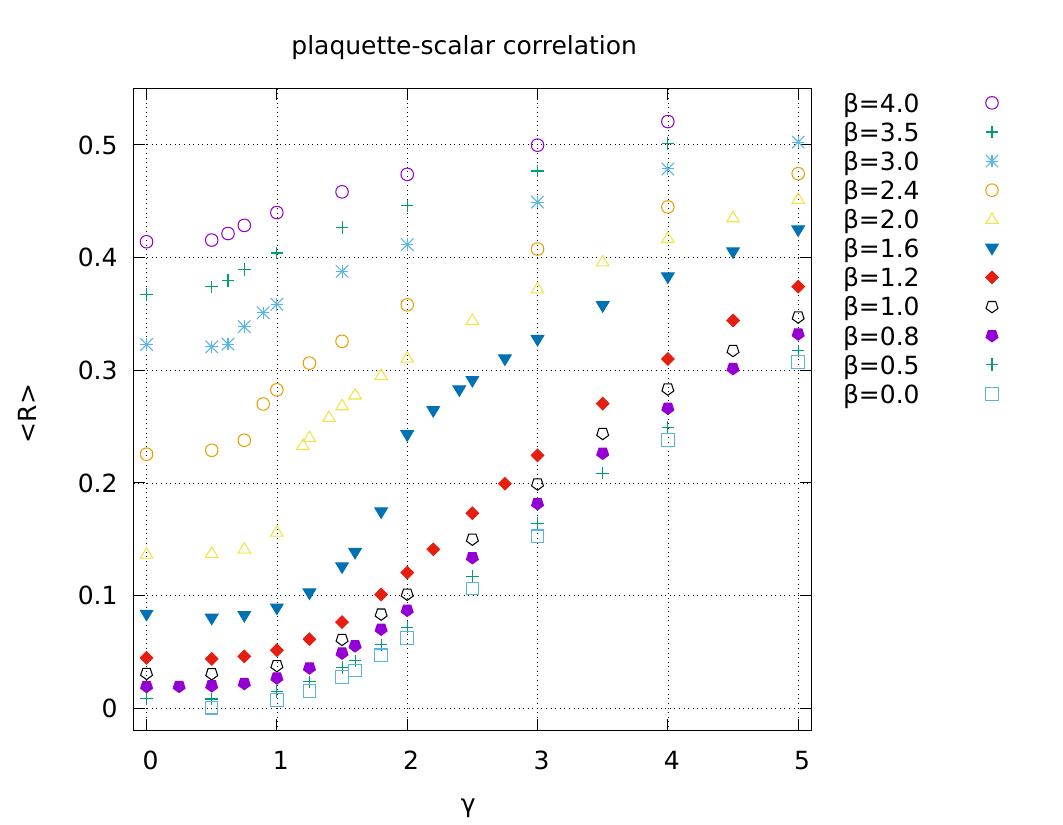}
	\includegraphics[height=60mm]{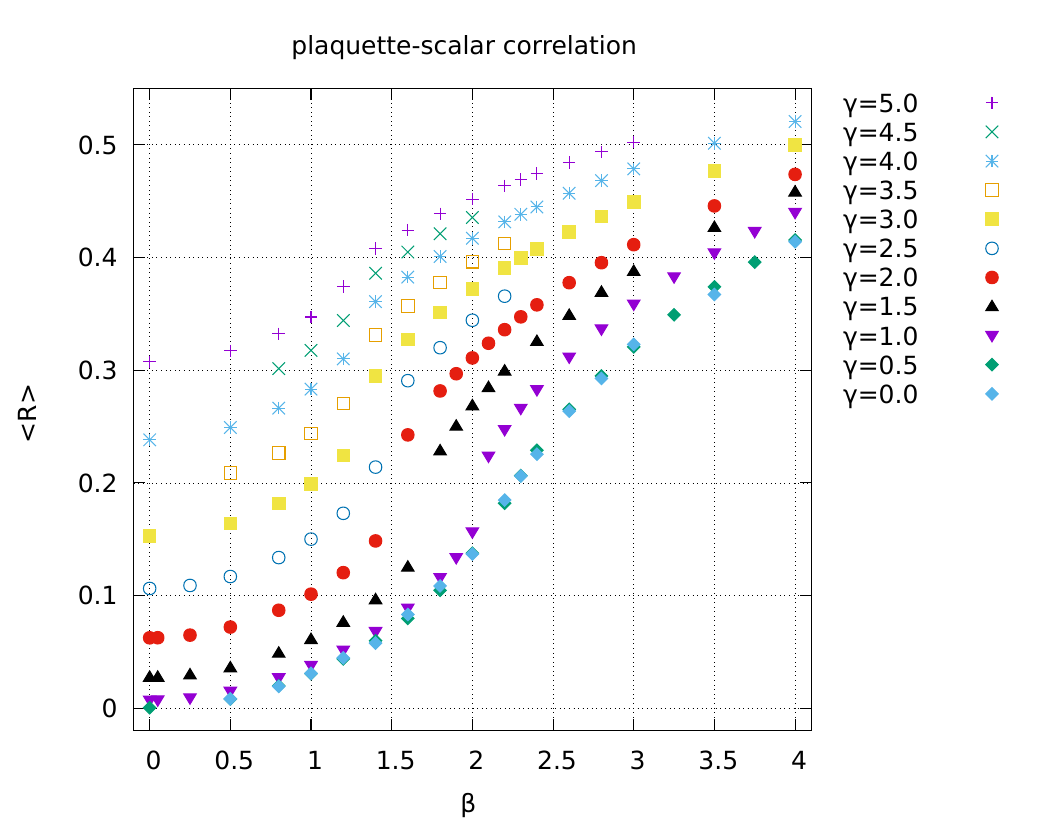}
	\caption{%
		Average of scalar-plaquette correration function  $\left\langle R \right\rangle$.
		Left: $\left\langle R \right\rangle$ versus $\gamma$ on various $\beta= \text{const.}$ lines. 
		Right: $\left\langle R \right\rangle$ versus $\beta$ on various $\gamma=\text{const.}$ lines. 
	}%
	\label{fig:measurement<R>}
\end{figure*}%
%%%%%%%%%%%%%%%%%
%%%%%%%
\begin{figure*}[hbt] \centering
\includegraphics[height=62mm]{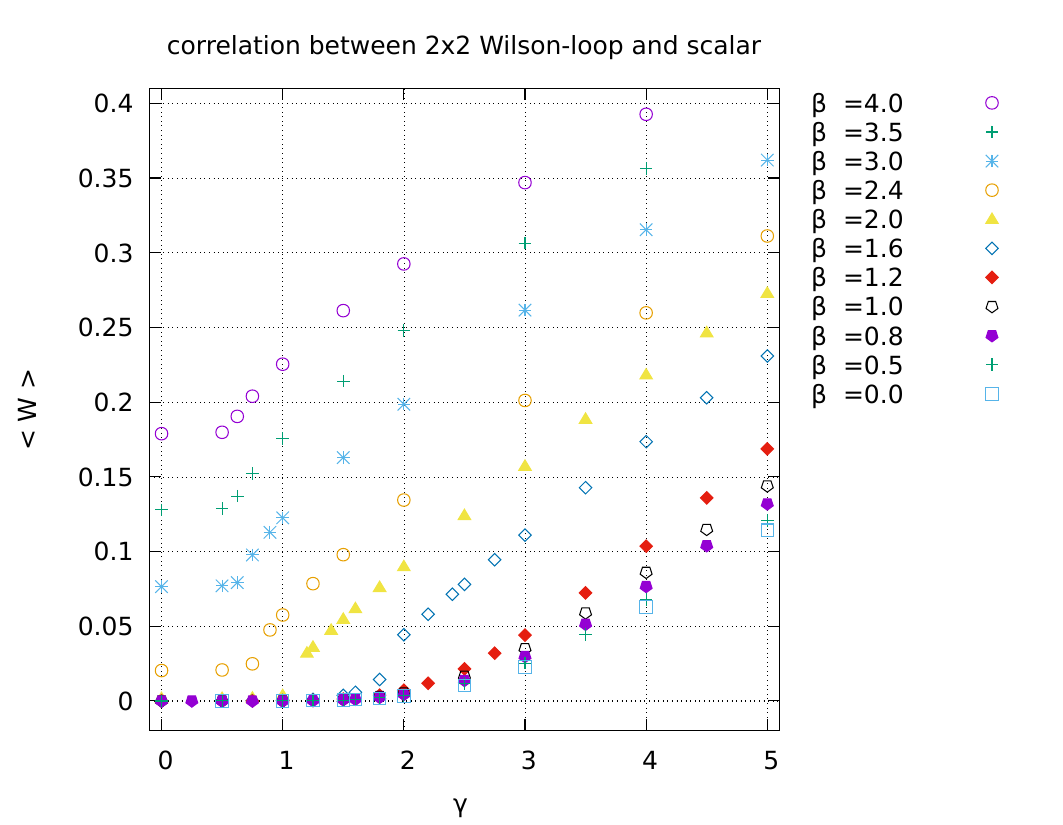} \quad
\includegraphics[height=62mm]{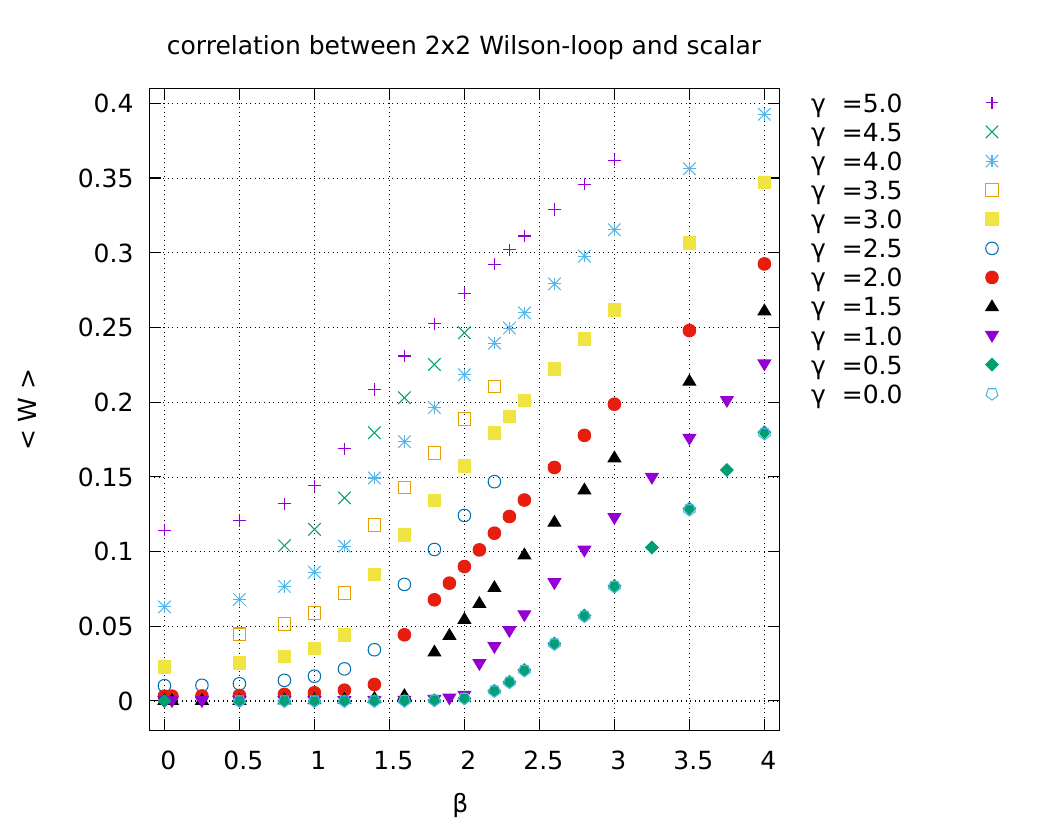}
\caption{%
Average of the correlation function  $\left\langle W \right\rangle$ between scalar and (2$\times$2)-Wilson loop.
Left: $\left\langle W \right\rangle$ versus $\gamma$ on various $\beta=\text{const.}$ lines. 
Right: $\left\langle W \right\rangle$ versus $\beta$ on various $\gamma=\text{const.}$ lines. 
}%
\label{fig:measurement<W>}%
\end{figure*}%
%%%%%%%%%%%%%%%%%

\subsection{Correlations between the scalar field and small Wilson loops}%

To confirm the results obtained in the previous section, we further examine two kinds of correlations 
between ($n \times n$)-Wilson loops $(n=1,2)$ and scalar fields.
They are defined in the gauge invariant way  using only the original field variables without relying on the color-direction field.

First, we consider the gauge-invariant correlation between a (1$\times$1)-Wilson loop (a plaquette) and scalar fields:
\begin{align}
	R &:=\frac{1}{6 N_{\text{site}}} \sum_{x,\mu<\nu}
	\frac{1}{2}\mathrm{tr}\left( 
    \overline{ U_{x,\mu\nu} \boldsymbol{\phi}_{x}U^{\dagger}_{x,\mu\nu} \boldsymbol{\phi}_{x} } 
	\right)  \,,	\label{eq:R}
\end{align}
where $\overline{ U_{x,\mu\nu} \boldsymbol{\phi}_{x}U^{\dagger}_{x,\mu\nu} \boldsymbol{\phi}_{x} } $ 
represents the averaged correlation defined by 
\begin{align}
  &	\overline{ U_{x,\mu\nu} \boldsymbol{\phi}_{x}U^{\dagger}_{x,\mu\nu} \boldsymbol{\phi}_{x} } \notag\\
  &:= \frac{1}{4}(    
      U_{x,\mu,\nu} \boldsymbol{\phi}_{x}U^{\dagger}_{x,\mu,\nu} \boldsymbol{\phi}_{x} 
     + U_{x,\nu,-\mu} \boldsymbol{\phi}_{x}U^{\dagger}_{x,\nu,-\mu} \boldsymbol{\phi}_{x} \notag \\
  &  +U_{x,-\mu,-\nu} \boldsymbol{\phi}_{x}U^{\dagger}_{x,-\mu,-\nu} \boldsymbol{\phi}_{x}
    +U_{x,-\nu,\mu} \boldsymbol{\phi}_{x}U^{\dagger}_{x,-\nu,\mu} \boldsymbol{\phi}_{x}
	   ),  \label{eq:hatR}
\end{align}
with  $U_{x,-\mu} = U^{\dagger}_{x-\hat{\mu}, \mu}$, 
which is identical to averaging the correlations between a scalar field and four plaquettes surrounding it
(see the left panel of Fig.\ref{fig:measurement-RW}). 
Here we want to define a lattice version of a local operator in the continuum measuring the correlation
between the gauge field strength and the scalar field by using plaquette variables 
and scalar field variables on the lattice. However, on the lattice such definitions are not unique.  
Therefore, to avoid such arbitrariness, we have defined the operator $R$ by averaging over the plaquettes around the scalar field to reduce the lattice artifact.

Next, we consider the gauge-invariant correlation between $(2\times 2)$-Wilson loops and scalar fields:
\begin{align}
	W &:=\frac{1}{6 N_{\text{site}}} \sum_{x,\mu<\nu}
	\frac{1}{2}\mathrm{tr}\left( 
	\widehat{W}^{(2\times 2)}_{x,\mu,\nu} \bm{\phi}_{x}\widehat{W}^{(2\times 2)\dagger}_{x,\mu,\nu} \bm{\phi}_{x} 
	\right) \label{eq:W}\,,
\end{align}
where we have introduced the variable $\widehat{W}^{(2\times 2)}_{x,\mu\nu}$ defined by the  (2$\times$2)-Wilson loop along a path $C$ (a 2$\times$2 rectangular loop) connected by the Schwinger lines 
(link variables):
\begin{align}
	\widehat{W}^{(2\times 2)}_{x,\mu\nu} := U_{x,\mu} 
	\left( \prod_{ <x,\rho> \in C} U_{x,\rho} \right) U_{x,\mu}^{\dagger} \, .  \label{eq:hatW}
\end{align}
%where the Wilson loop along the path $C$ (2$\times$2 rectangular) is connected by
Here we have an arbitrariness in defining the correlation between the Wilson loop and the scalar field, at which site the scalar field is located. 
We here define the correlation between the Wilson loop and the scalar field at the center $x$ of the Wilson loop so that $W$ has similar information to $R$. To construct a gauge-invariant operator, we connect the Wilson loop and the scalar field $\boldsymbol{\phi}_{x}$ at the center using the Schwinger lines (see the right panel of Fig.\ref{fig:measurement-RW}).

Now we examine the data for the correlation between the scalar field and Wilson loop: 
 $\left\langle R \right\rangle$   and  $\left\langle W \right\rangle$. 
The phase boundaries are searched in the same way as in the previous section.

Figure \ref{fig:measurement<R>} shows the measurements of $\left\langle R \right\rangle$. 
The left panel shows plots of $\left\langle R \right\rangle $ versus $\gamma$ along various $\beta=\text{const.}$ lines. 
The right panel shows plots of $\left\langle R \right\rangle $ versus $\beta$ along various $\gamma=\text{const.}$ lines.

Figure \ref{fig:measurement<W>} shows the measurements of $\left\langle W \right\rangle$. 
The left panel shows plots of $\left\langle W \right\rangle $ versus $\gamma$ along various $\beta=\text{const.}$ lines. 
The right panel shows plots of $\left\langle W \right\rangle $ versus $\beta$ along various $\gamma=\text{const.}$ lines.

\begin{figure*}[thb] \centering
	\includegraphics[height=63mm] {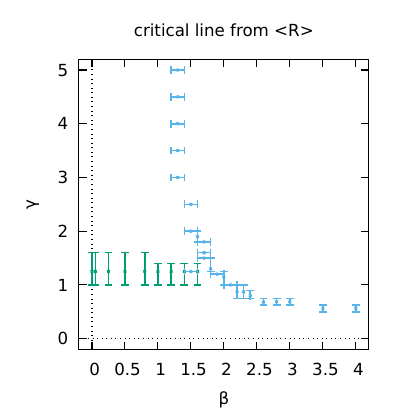} \quad \quad
	\includegraphics[height=63mm] {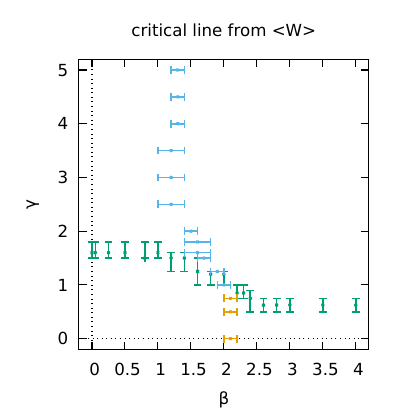}
	\caption{
	Critical lines determined (left) from $\left\langle R \right\rangle$, and (right) from $\left\langle W \right\rangle$.
	}%
	\label{fig:critical<R><W>}
\end{figure*}

The left panel of Fig.\ref{fig:critical<R><W>} shows the phase boundary determined  from  $\left\langle R \right\rangle$. 
The blue boundary ($\beta_c(\gamma)$) corresponds to the location at which $\left\langle R \right\rangle$ has gaps or bends. 
The green boundary ($\gamma_c(\beta)$) corresponds to the location 
at which the function $\left\langle R \right\rangle$ changes shape from a constant function $\left\langle R \right\rangle = \text{const.} $ 
to the increasing function.
These phase boundaries reproduce those determined from the scalar-color correlation, i.e., 
$\left\langle \left\vert Q\right\vert \right\rangle$ and $\left\langle Q^{2}\right\rangle$
(see Fig.\ref{fig:critical<Q>}).

%%%%%%%%%%%%%%%%%%%%%
%%%%%%%%%%%%%%%%%%%%%%%

The right panel of Fig.\ref{fig:critical<R><W>} shows the phase boundary determined 
from $\left\langle W \right\rangle $. 
The blue boundary ($\gamma_c(\beta)$) corresponds to the location at which $\left\langle W \right\rangle$ has gaps or bends. 
The green ($\gamma_c(\beta)$) and orange ($\beta^{\prime}_c(\gamma)$) boundaries correspond to locations 
at which the function $\left\langle W \right\rangle$ changes the shape from a constant function to the increasing function. 
The green boundary shows the function $\left\langle W \right\rangle (\beta)$ jumps at the critical points. 
When we focus on blue and green boundaries ($\beta_c(\gamma)$ and $\gamma_c(\beta)$),  
the result reproduces the phase diagram determined from the scalar-color correlation (see Fig.\ref{fig:critical<Q>}).

On the other hand, the orange boundary seems to separate the region below the green boundary into two regions:
in the region $\beta < \beta^{\prime}_c(\gamma)$ we have no correlation between the Wilson loop and the scalar field, $\left\langle W \right\rangle \simeq 0$,  while 
in the region   $\beta > \beta^{\prime}_c(\gamma)$  $\left\langle W \right\rangle$ takes nonzero value, $\left\langle W \right\rangle \simeq >0 $.
However,  the orange boundary shows 
the location at which $\left\langle W \right\rangle)$
has the bent, although it is continuous without any gaps.
On the other hand, we cannot find such a boundary from the measurement of $\left\langle R \right\rangle (\beta)$ 
(in sharp contrast to that of  $\left\langle W \right\rangle$.)
Therefore, this orange boundary could be the crossover
as was discussed in Sec.\ref{sec:sus_P&M}.

In Fig. 16, although 
the transition line from  $\left\langle W \right\rangle$ is located a little higher than the transition line from $\left\langle R \right\rangle$,  
these two transition lines agree within errors of the numerical simulations, 
as seen by the error bars. % in Fig.16.  

\section{Understanding the new phase structure obtained from numerical simulations}

\begin{figure}[hbt]
\centering
\includegraphics[height=65mm] {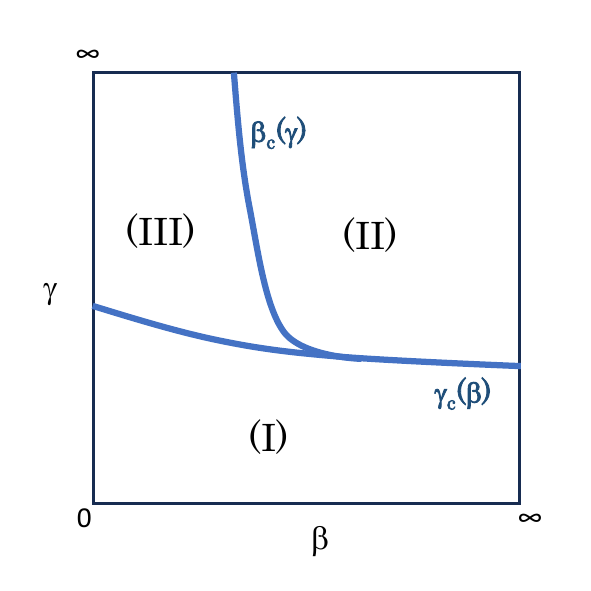}
\caption{%
Phase structure of the lattice SU(2) gauge-adjoint scalar model: %
(I) confinement phase, (II)  Higgs phase, (III)  confinement phase. }
\label{fig:phase_diagram}
\end{figure}

Finally, we discuss why the above phase structure should be obtained and how the respective phase is characterized from the physical point of view. Figure \ref{fig:phase_diagram} shows the schematic view of the resulting phase structure.

(i) 
First, we consider the region below the new critical line $\gamma<\gamma_c(\beta)$.
In the limit $\gamma \to 0$, especially, the $SU(2)$ gauge-scalar model reduces to the pure compact $SU(2)$ gauge model which is expected to have a single confinement phase with no phase transition and has a mass gap on the whole $\beta$ axis in four spacetime dimensions \cite{Creutz82}. 
Confinement is expected to occur due to vacuum condensations of ``non-Abelian'' magnetic monopoles \cite{dualsuper}.  Here the ``non-Abelian'' magnetic monopole should be carefully defined gauge independently using the gauge-invariant method, which is actually realized by extending the gauge-covariant decomposition of the gauge field; see \cite{KKSS15} for a review.  
More comments will be given below. 

Even in the region $0 < \gamma<\gamma_c(\beta)$, the effect of the scalar field would be relatively small and confinement would occur  in the way similar to the pure $SU(2)$ gauge theory, which we call confinement phase (I).
Confinement phase (I)  is regarded as a disordered phase in the sense that the color direction field $\bm{n}_{x}$ takes various possible directions with no specific direction in color space.
This can be estimated through $Q$ in relation to the direction of the adjoint scalar field $\bm{\phi}_{x}$, which yields very small or vanishing values of the average
 $\left\langle \left\vert Q\right\vert \right\rangle=0$.

(ii)
Next, we consider the region above the new critical line $\gamma>\gamma_c(\beta)$ where 
$\left\langle \left\vert Q\right\vert \right\rangle$
 takes the nonvanishing value 
$\left\langle \left\vert Q\right\vert \right\rangle>0$,
including the two phases: Higgs phase (II) $\gamma>\gamma_c(\beta)$, $\beta>\beta_c(\gamma)$ and confinement phase (III) $\gamma>\gamma_c(\beta)$, $\beta<\beta_c(\gamma)$.
In order to consider the difference between the two phases (II) and (III), we first  consider the limit $\gamma \to \infty$.  
In this limit, the $SU(2)$ gauge-scalar model reduces to the pure compact $U(1)$ gauge model. 
The pure compact $U(1)$ gauge model in four spacetime dimensions has two phases: confinement phase with massive $U(1)$ gauge field in the strong gauge coupling region $\beta<\beta_*=\beta_c(\infty)$ and the Coulomb phase with massless $U(1)$ gauge field in the weak gauge coupling region $\beta>\beta_*=\beta_c(\infty)$, which has been proved rigorously \cite{Guth80,FS82}. 
Confinement in the compact $U(1)$ gauge model in the strong gauge coupling region $\beta<\beta_*=\beta_c(\infty)$ is understood based on the $U(1)$ magnetic monopole as shown in \cite{BMK77}. 

For large but finite $\gamma < \infty$, furthermore, the critical line $\beta=\beta_c(\gamma)$ extends into the interior of the phase diagram from the critical point $(\beta_*=\beta_c(\infty),\gamma=\infty)$ as shown in \cite{Brower82} by integrating out the scalar field to obtain the effective gauge theory, which supports the above two regions even for finite $\gamma < \infty$. 

In the  Higgs phase (II), $\beta>\beta_c(\gamma)$ above the new critical line $\gamma>\gamma_c(\beta)$ with a finite $\gamma < \infty$, the off diagonal gauge fields for the modes $SU(2)/U(1)$ become massive due to the BEH mechanism, which is a consequence of the (partial) spontaneous symmetry breaking $SU(2) \to U(1)$ according to the conventional understanding of the BEH mechanism, although this phenomenon is also understood gauge independently based on the new understanding of the  BEH mechanism without the spontaneous symmetry breaking \cite{Kondo16}. 
Therefore, the diagonal gauge field  for the mode $U(1)$ always remains massless everywhere in the phase (II).  
This is not the case in the other phases. 
Therefore, the Higgs phase (II) can be clearly distinguished from the other phases. 
In the limit $\gamma \to \infty$, especially, the off diagonal gauge fields become infinitely heavy and decouple from the theory, while the diagonal gauge field survives the limit both in (II) and (III). 
Consequently, the $SU(2)$ gauge-scalar model reduces to the pure compact $U(1)$ gauge model. 
This observation is consistent with the above consideration in the limit $\gamma \to \infty$.
%for the right region (II):$\beta>\beta_c(\gamma)$ Higgs phase (II) for the existence of the massless $U(1)$ gauge field. 

%the correlator between the original adjoint scalar field $\bm{\phi}_{x}$ and the color-direction field $\bm{n}_{x}$ take the non-vanishing value $\langle Q \rangle>0$.  
The nonvanishing value  $\langle \vert Q \vert \rangle>0$  means that the color-direction field $\bm{n}_{x}$ correlates strongly with the given scalar field $\bm{\phi}_{x}$ which tends to align to an arbitrary but a specific direction in the region $\gamma>\gamma_c(\beta)$ as expected from the spontaneous symmetry breaking 
in an ordered phase.

(iii)
In the left  region (III) $\beta<\beta_c(\gamma)$ above the new critical line $\gamma>\gamma_c(\beta)$ to be identified with another confinement phase (III), the gauge fields become massive due to different physical origins.  
In the region (III), indeed, the gauge fields become massive due to self-interactions among the gauge fields, as in the phase (I).  
In the confinement phase (III), no massless gauge field exists and the gauge fields for all the modes become massive, which is consistent with the belief that the original gauge symmetry $SU(2)$ would be kept intact and not spontaneously broken.

In the confinement phases (I) and (III)  there occur magnetic monopole condensations which play the dominant role in explaining quark confinement based on the dual superconductor picture, while in the Higgs phase (II)  there are no magnetic monopole condensations and confinement would not occur. 
However, it should be remarked that the origin of magnetic monopoles is different in the two regions, (I) and (III). 
In (III) the magnetic monopole is mainly originated from the adjoint scalar field just like the `t Hooft-Polyakov magnetic monopole in the Georgi-Glashow model \cite{Polyakov77}. 
In (I)  the magnetic monopole is constructed from the gauge field.  Indeed, the magnetic monopole can be constructed only from the gauge degrees of freedom, which is explicitly constructed from the color direction field  in the gauge-invariant way \cite{KKSS15}.

\section{Conclusion and discussion}

In this paper, we have investigated the lattice $SU(2)$ gauge-scalar model
with the scalar field in the adjoint representation of the gauge group in a
gauge-independent way.
%We first explain the lattice construction of the gauge-independent description of the BEH mechanism, which does not rely on the spontaneous breaking of gauge symmetry.
This model was considered to have two completely separated confinement and Higgs phases according to the preceding studies \cite{Brower82} based on numerical simulations which
have been performed in the specific gauge fixing based on the conventional
understanding of the Brout-Englert-Higgs mechanism \cite{Higgs1}.

We have reexamined this phase structure in the gauge-independent way based on
the numerical simulations performed without any gauge fixing, which should be compared with  the preceding studies \cite{Brower82}. This is motivated to confirm the recently
proposed gauge-independent Brout-Englert-Higgs mechanics for giving the mass of the
gauge field without relying on any spontaneous symmetry breaking \cite{Kondo16,Kondo18}. For this purpose we have investigated correlation  
between gauge-invariant operators
obtained by combining the original adjoint scalar field and the new field
called the color-direction field which is constructed from the gauge field
based on the gauge-covariant decomposition of the gauge field due to
Cho-Duan-Ge-Shabanov and Faddeev-Niemi. 

Consequently, we have reproduced gauge independently the transition line separating
confinement and Higgs phase obtained in \cite{Brower82}, and show surprisingly
the existence of a new transition line that divides completely the confinement
phase into two parts. 
Moreover, we have shown that the same new transition line as the above is also obtained based on the gauge-invariant operators defined by the original gauge field and scalar fields alone without using the color-direction field. 

We have discussed the physical meaning of the new transition and implications to confinement mechanism. 
To establish our claim given in this paper, it is important to show how the distinction between (I) and (III) manifests itself in physical observables, such as the string tension, mass spectrum, or scattering amplitudes. 
In order to continue our research to obtain the evidence in this direction, we plan to measure the mass spectrum and magnetic monopole condensation in each phase to examine our claim, in addition to some theoretical considerations possibly derived from the analytical calculations in  subsequent papers, together with more discussions on the physical properties of the respective phase.
In particular, it is quite important to study whether or not the new phase (III) is  a lattice artifact  and survives the continuum limit. 

The result obtained in this paper should be compared with the lattice $SU(2)$
gauge-scalar model with the scalar field in the fundamental representation of
the gauge group in a gauge-independent way. This model has a single
confinement-Higgs phase where two confinement and Higgs regions are
analytically continued according to the preceding studies \cite{Ostewlder78,FradkinShenker79}. Even in this case, it is shown \cite{Ikeda23} that 
the composite operator constructed from the original fundamental scalar field
and the color-direction field can discriminate two regions and indicate the
existence of the transition line separating the confinement-Higgs phase into
two completely different phases, the confinement phase and Higgs phase.

\section*{ACKNOWLEDGMENT}
This work was supported by Grant-in-Aid for Scientific Research, JSPS KAKENHI
Grant No. (C) No.23K03406. The numerical simulation is supported by the
Particle, Nuclear and Astro Physics Simulation Program No.2022-005 (FY2022) of
Institute of Particle and Nuclear Studies, High Energy Accelerator Research
Organization (KEK). 
This research was supported in part by the Multidisciplinary Cooperative Research Program
in CCS, University of Tsukuba. 
The authors would like to thank Professor Jun Nishimura for fruitful discussions.

%%%%%%%%%%%%%%%%%%%%%%%% Appendix %%%%%%%%%%%%%%%%%%%%%%%%%%%%%%%%%%%%
\appendix

%%%%%%%%%%%%%%%%% figure for Appendix %%%%%%%%%%%%%%%

\begin{figure*}[bht] \centering
\includegraphics[height=60mm] {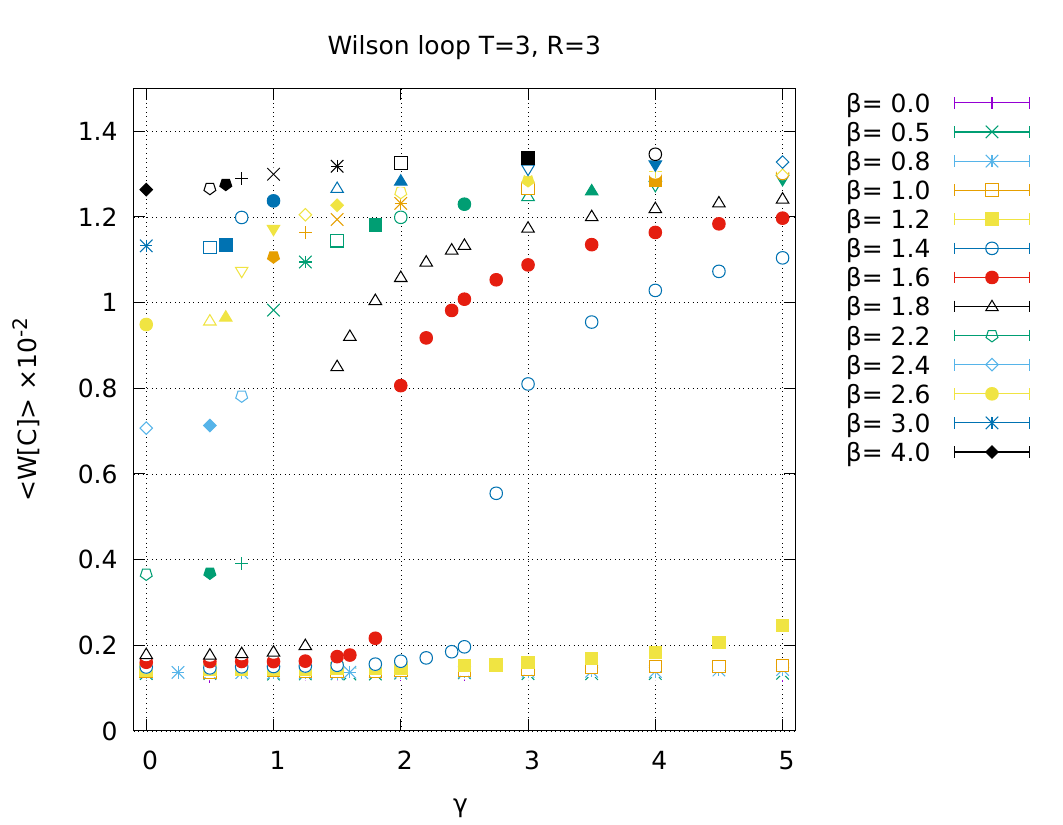} \hspace{5mm}
\includegraphics[height=60mm] {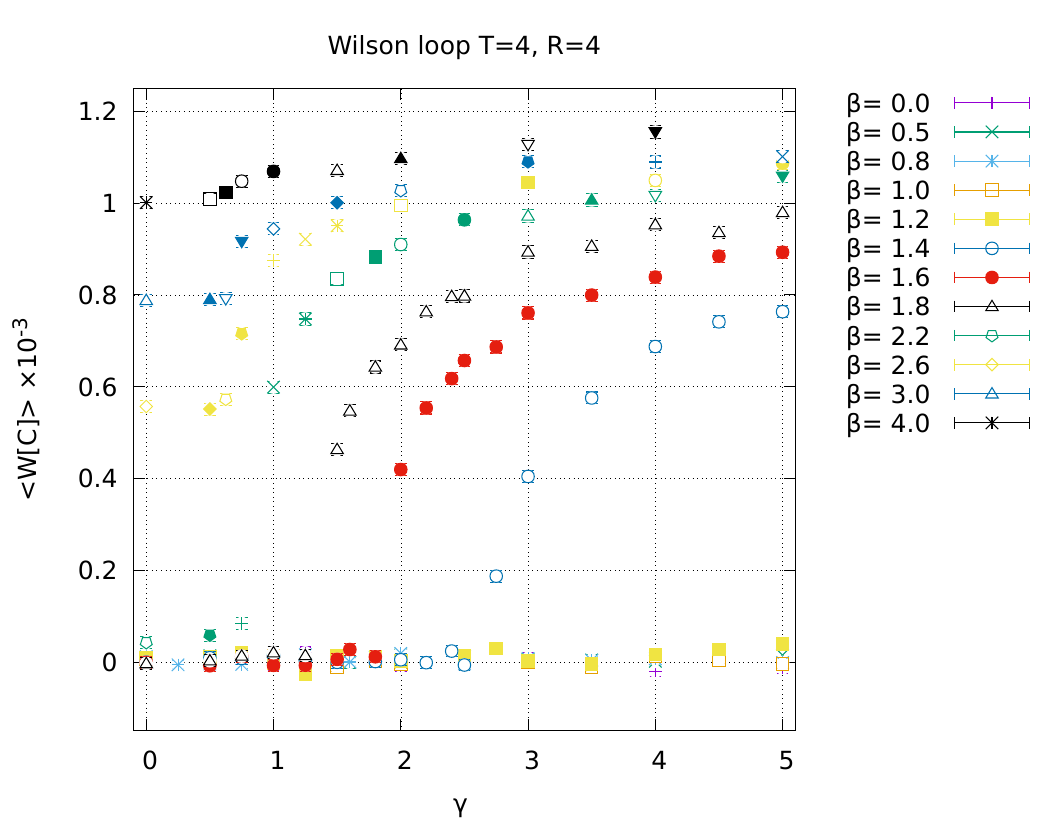}
\caption{
Average of the Wilson loops versus $\gamma$ on various $\beta=\text{const.}$ lines. 
Left:  $\left\langle W(3,3) \right\rangle$.
Right: $\left\langle W(4,4) \right\rangle$.
}
\label{fig:measure<Wc>}%
\end{figure*}%

%%%%%%%%%%

\begin{figure*}[bht] \centering
\includegraphics[height=60mm] {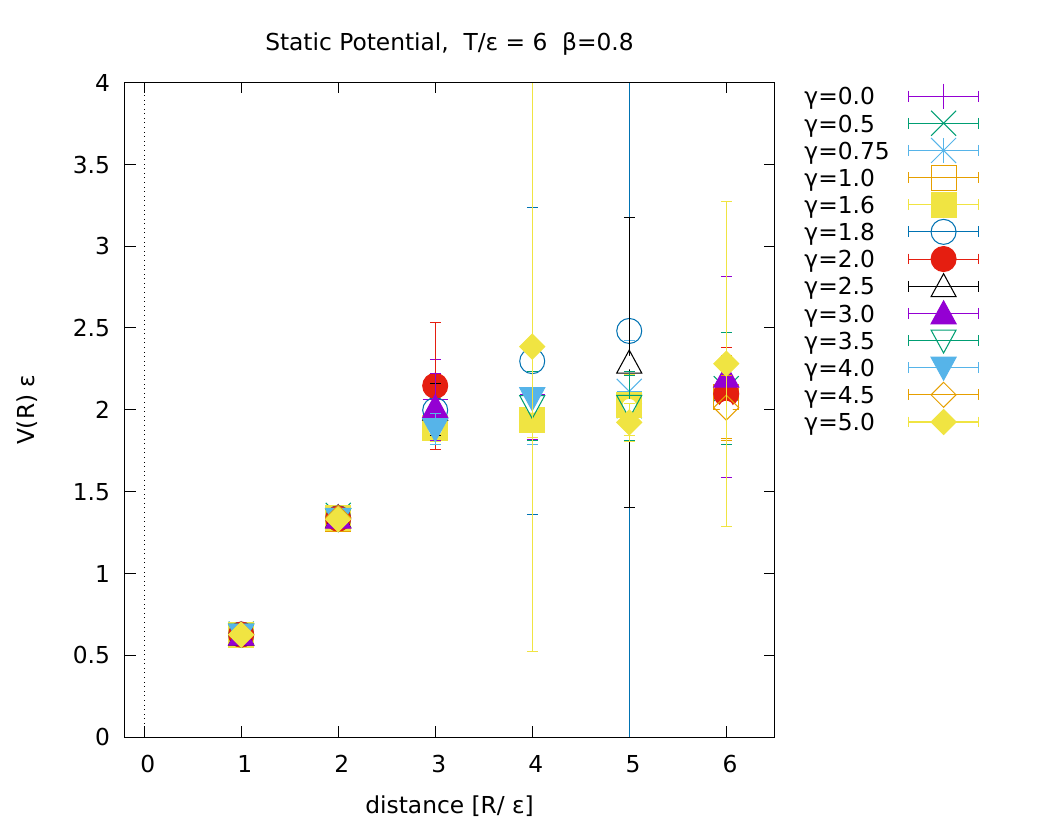} \hspace{5mm}	\includegraphics[height=60mm] {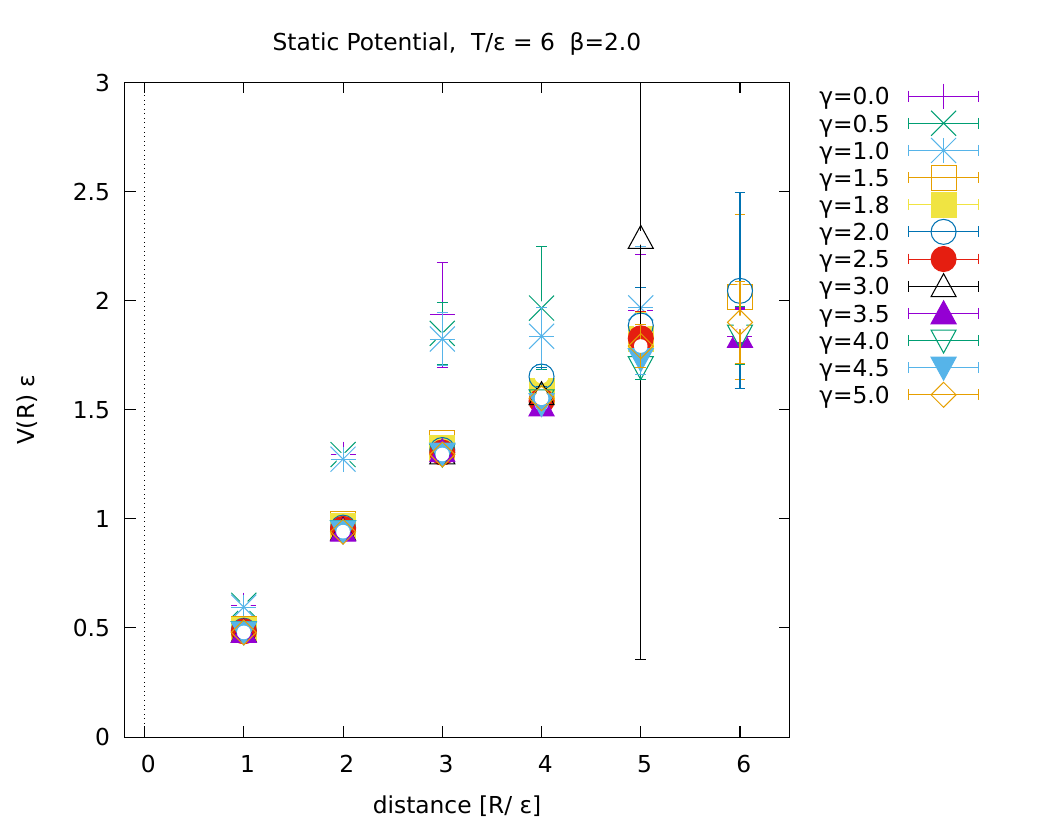}
\caption{
Static potential $V(R,T/\epsilon=6)$ obtained from the Wilson-loop average for various $\gamma$ along the $\beta=\text{const.}$ line. 
Left: $\beta=0.8$. Right: $\beta=2.0$.
}
\label{fig:potential}%
\end{figure*}%

%%%%%%%%%%%%%%%%%%%%%%%%%%%%%%%%%%%%%%%%%%

\section{Phase boundary examined by the Wilson-loop average}

To understand the nature of the new transition between the two confinement phases (I) and (III), we further examine the Wilson-loop operator defined by
\begin{align}
	W^{(R \times T)}_{C} := \frac{1}{2}\text{tr} \left[\prod_{ <x,\rho> \in C } U_{x,\rho} \right ] \, , %\notag \\
	% W(R, T) &:= \frac{1}{6} \frac{1}{N_\text{site}}
	% \sum_{ \{C\} } W^{(R \times T)}_{C} \, ,
\end{align} 
where the path $C$ represents the loop consisting of links $<x,\rho>$ on the perimeter of a ($R \times T$) rectangle. 
%and the sum $\sum_{ \{C\} }$ runs over all possible ($R \times T$) rectangles on the planes specified by $\mu$ and $\nu$ labelling the directions of a plane on the four-dimensional lattice. 
%in the $\mu$-$\nu$ plain with $\mu$ and $\nu$ being labels of the four directions.
To reduce the statistical errors, we use the averaged operator $W(R, T)$ reflecting the translation and rotation invariance of the Wilson-loop operator:
%over all possible Wilson loops $\{ C \}$.
\begin{align}
	%W^{(R \times T)}_{C} &:= \frac{1}{2}\text{tr} \left[\prod_{ <x,\rho> \in C } U_{x,\rho} \right \} \, , \notag \\
	W(R, T) := \frac{1}{N_\text{site}} \frac{1}{6} 
	\sum_{ \{C\} } W^{(R \times T)}_{C} \, ,
\end{align} 
where the sum $\sum_{ \{C\} }$ runs over all possible ($R \times T$) rectangular loops $\{ C \}$ on the six planes on the four-dimensional lattice. 
%specified by $\mu$ and $\nu$ labelling the directions of a plane on the four-dimensional lattice. 
In addition, we apply the APE smearing method \cite{APE} to reduce the noise. 

We first measure a Wilson-loop average $\left\langle W(R,T)\right\rangle$.
The search for a phase boundary is performed by measuring 
the Wilson-loop average $\left\langle W(R,T)\right\rangle$ by changing $\gamma$ along the $\beta= \text{const.}$ lines.
To identify the phase boundary, we use the bent, step, and gap 
observed in the graph for $\left\langle W(R,T)\right\rangle$
as is done for $\left\langle P \right\rangle$.

Figure \ref{fig:measure<Wc>} shows the results of measurements of the Wilson-loop average $\left\langle W(R,T)\right\rangle$ in the $\beta$--$\gamma$ plane. 
The left panel of Fig.\ref{fig:measure<Wc>} shows plots of 
$\left\langle W(3,3)\right\rangle $ versus $\gamma$ along the $\beta=\text{const.}$ lines. 
For smaller values of $\beta$: $\beta < 1.2$, $\left\langle W(3,3)\right\rangle $
takes almost the same constant value for changing $\gamma$.
We find no phase boundary between the phases (I) and (III) from measurements of the Wilson-loop average.
For larger values of $\beta$: $\beta \ge 1.2$, on the other hand, 
$\left\langle W(3,3)\right\rangle $ is constant up to a critical point
$\gamma_{c}(\beta)$ as $\gamma$ increases, 
but for $\gamma>\gamma_{c}(\beta)$ it changes rapidly to take larger values.
Thus we identify the critical line $\gamma_{c}(\beta)$ with the phase boundary from the Wilson-loop average. 
Moreover, the right panel of Fig.\ref{fig:measure<Wc>} shows plots of 
$\left\langle W(4,4)\right\rangle $ versus $\gamma$ along the $\beta=\text{const.}$ lines. 
We obtain the same results on the phase boundary as in the case of 
$\left\langle W(3,3)\right\rangle $. 
Note that these results are consistent with that obtained by 
$\left\langle P \right\rangle$ 
(see the left panel of Fig.\ref{fig:mesurement<P>}).

Next, we examine the static potential from the Wilson-loop average 
to obtain the information about the string tension:
\begin{align}
	V(R,T) = - \frac{1}{T} \log \left\langle W(R,T) \right\rangle \, .
	\label{eq:potential}
\end{align}
Figure \ref{fig:potential} shows the plots of static potentials for various values of $\gamma$ along the $\beta=\text{const.}$ lines. The left panel of Fig.\ref{fig:potential} shows the plots of the static potentials,%
\footnote{
	%In the measurement of 
	For smaller values of $\beta$, the Wilson-loop averages have smaller values, but the errors are large compared with those for larger values of $\beta$. 
	Therefore, as for $\beta=0.8$, we use additional 15,000 configurations for measurements to obtain signals.%
} %% end of footnote %%% 
$V(R, T/\epsilon =6)$, for various values of $\gamma$ along the $\beta=0.8$ line which crosses the phase boundary between the phases (I) and (III).
The potential $V$ increases as $R$ increases up to the critical distance $R_c
\simeq 3$ and immediately becomes flat for $R>R_c$, which could be the string breaking 
due to the screening of the gauge field by the pair creation of the scalar particles from the vacuum. 
All plots of the static potential overlap within errors irrespective of the values of $\gamma$. 

The right panel of Fig.\ref{fig:potential} shows 
the plots of the static potentials, $V(R, T/\epsilon =6)$, for various values of $\gamma$ along the $\beta=2.0$ line which crosses the phase boundary between the phases (I) and (II).
We find two kinds of potentials depending on the value of $\gamma$: smaller or larger than a critical value $\gamma_c(\beta)$. 
In the confinement phase (I) ($\gamma < \gamma_c(\beta)$) the potential $V$ increases as $R$ increases up to the critical distance $R_c\simeq 3$ and becomes flat for $R>R_c$,
while in the Higgs phase (II) ($\gamma > \gamma_c(\beta)$) the potential $V(R)$ increases as $R$ increases in the whole range of $R$, which could be identified with the Yukawa potential.

In the text, we have shown that the phase boundary separating confinement phases (I) and (III) can be elucidated through the correlation between the scalar field and the Wilson loop (or the color direction field extracted by the gauge-covariant decomposition). 
As shown in this Appendix, indeed, we cannot detect any phase boundary that separates confinement phases (I) and (III) by using the Wilson-loop operator alone without taking into account its correlation with the scalar field.
This result shows that the new transition we claim is not a transition detectable by the string tension.

%%%%%%%%%%%%
\vfill
\pagebreak


\begin{thebibliography}{99}                                                                                               %
\bibitem {Higgs1}P.W. Higgs,
%Broken symmetries, massless particles and gauge fields,
Phys. Lett. \textbf{12}, 132 (1964)%
;
%\\
%P.W. Higgs,
%Broken Symmetries and the Masses of Gauge Bosons,
Phys. Rev. Lett. \textbf{13}, 508 (1964).
%--509
%\\
%P.W. Higgs,
%Spontaneous Symmetry Breakdown without Massless Bosons
%Phys. Rev. \textit{145}, 1156--1163 (1966).

%\bibitem{Higgs2}
F. Englert and R. Brout,
%Broken Symmetry and the Mass of Gauge Vector Mesons,
Phys. Rev. Lett.\textbf{13}, 321
%--323
(1964).

\bibitem {Brower82}R.C. Brower, D.A. Kessler, T. Schalk, H. Levine, 
and M. Nauenberg, Phys. Rev. D\textbf{25}, 3319 (1982).

\bibitem {Elitzur75}S. Elitzur,
%Impossibility of spontaneously breaking local symmetries,
Phys. Rev. D\textbf{12}, 3978 (1975).

\bibitem {Kondo16}K.-I. Kondo,
%Gauge-invariant description of Higgs phenomenon and quark confinement,
Phys. Lett. B \textbf{762}, 219
%--224
(2016).
%CHIBA-EP-219
%DOI: 10.1016/j.physletb.2016.09.026
arXiv:1606.06194 [hep-th]

\bibitem {Kondo18}K.-I. Kondo,
%Gauge-independent Brout-Englert-Higgs mechanism and Yang-Mills theory with a gauge-invariant gluon mass term,
%CHIBA-EP-230,
Eur. Phys. J. C \textbf{78}, 577 (2018). arXiv:1804.03279 [hep-th]

\bibitem {Cho80}Y.M. Cho, Phys. Rev. D\textbf{21}, 1080 (1980)%.
;  Phys. Rev.D \textbf{23}, 2415 (1981).

\bibitem {Duan-Ge79}Y.S. Duan and M.L. Ge, Sinica Sci. \textbf{11}, 1072
%--1081
(1979).

\bibitem {Shabanov99}S.V. Shabanov, Phys. Lett. B\textbf{463}, 263
%--272
(1999). [hep-th/9907182]

%S.V. Shabanov, Phys. Lett. B\textbf{458}, 322 %--330
%(1999). [hep-th/0608111];

\bibitem {FN98}L.D. Faddeev and A.J. Niemi, Phys. Rev. Lett. \textbf{82}, 1624
(1999). [hep-th/9807069];

L.D. Faddeev and A.J. Niemi, Nucl. Phys. B\textbf{776}, 38
%--65
(2007). [hep-th/0608111]

\bibitem {Exactdecomp09}A. Shibata, K.-I.Kondo, T.Shinohara, Phys.
Lett.B\textbf{691}, 91
%--98
(2010). arXiv:0706.2529 [hep-lat]

\bibitem {CFNdccomp07}A. Shibata, S. Kato, K.-I. Kondo, T. Murakami, T.
Shinohara, and S. Ito,
%Compact lattice formulation of Cho-Faddeev-Niemi decomposition: gluon mass generation and infrared Abelian dominance
Phys. Lett. B\textbf{653}, 101
%--108
(2007). arXiv:0706.2529 [hep-lat]

\bibitem {KKSS15}K.-I. Kondo, S. Kato, A. Shibata and T. Shinohara, Phys.
Rept. \textbf{579}, 1--226 (2015). arXiv:1409.1599 [hep-th]

\bibitem {dualsuper}
%\bibitem{Nambu74}
Y. Nambu,
%Strings, monopoles, and gauge fields,
Phys. Rev. D\textbf{10}, 4262 (1974). \newline G. 't Hooft, 
in {\it  High Energy Physics}, edited by A. Zichichi (Editorice Compositori, Bologna, 1975).
\newline S. Mandelstam,
%Vortices and quark confinement in non-abelian gauge theories,
Phys. Rep. \textbf{23}, 245 (1976).

%\bibitem {tHooft81}G. 't Hooft,
%Topology of the gauge condition and new confinement phases in non-Abelian gauge theories,
%Nucl. Phys. B\textbf{190} [FS3], 455 (1981).

%\bibitem {EI82}Z.F. Ezawa and A. Iwazaki,
%Abelian dominance and quark confinement in Yang-Mills theories,
%Phys. Rev. D\textbf{25}, 2681 (1982).


%\bibitem {AS99}K. Amemiya and H. Suganuma,
%Off diagonal gluon mass generation and infrared Abelian dominance in the maximally Abelian gauge in lattice QCD,
%Phys. Rev. D\textbf{60}, 114509 (1999). [hep-lat/9811035]


\bibitem {BhanotCreutz81}G. Bhanot and M. Creutz Phys. Rev. D\textbf{24}, 3212 (1981).


\bibitem {Creutz82} 
M. Creutz, 
{ \it Quarks, Gluons and Lattices (Cambridge Monographs on Mathematical Physics }
(Cambridge University Press, 1985).


\bibitem {Guth80} A.H. Guth, Phys. Rev. D\textbf{21}, 2291 (1980).


\bibitem {FS82} J. Fröhlich and T. Spencer, Commun. Math. Phys. \textbf{83}, 411 %-–454 
(1982).

\bibitem{BMK77} T. Banks, R. Myerson, and J.B. Kogut, Nucl. Phys. B\textbf{129},  493 (1977).

\bibitem {Polyakov77} A.M. Polyakov,  
%Quark confinement and topology of gauge theories,
Nucl. Phys. {\bf B120}, 429 %--458
 (1977).


\bibitem {Ostewlder78}K. Ostewalder and E. Seiler, Annls. Phys \textbf{110},
440 (1978).

\bibitem {FradkinShenker79}E. Fradkin and S.H. Shenker, Phys. Rev.
D\textbf{19}, 3682 (1979).

%\bibitem {PhysRep15}K.-I. Kondo, S. Kato, A.Shibata, T. Shinohara, Phys. Rept. \textbf{579}, 1--226 (2015) . e-Print:1409.1599 [hep-th]

\bibitem {Ikeda23} R. Ikeda, S. Kato, K.-I. Kondo, and A. Shibata,  Phys. Rev. D 109, 054505 (2024), arXiv:2308.13430, CHIBA-EP-259, KEK Preprint 2023-27

%\bibitem {SY90}T. Suzuki and I. Yotsuyanagi,
%Possible evidence of abelian dominance in quark confinement,
%Phys. Rev. D\textbf{42}, 4257 (1990).

\bibitem{APE} M. Albanese et al. (APE Collaboration), Phys. Lett. B\textbf{192}, 163–169 (1987).

\end{thebibliography}
\end{document}